\documentclass[10pt,aps,pre,twocolumn,nofootinbib,floatfix
]
{revtex4-1}

\usepackage{amssymb,amsmath}
\usepackage{graphicx}
\usepackage[usenames,dvipsnames,svgnames,table]{xcolor}
\usepackage[normalem]{ulem}
\usepackage{hyperref}
\usepackage{siunitx}
\usepackage{bm}
\usepackage{gensymb}
\usepackage{placeins}
\usepackage[caption=false]{subfig}
\hypersetup{
	pdfauthor={},     		 
	colorlinks=true,       	 
	linkcolor=blue,          
	citecolor=red           
}

\allowdisplaybreaks

\newcommand\md{\mathrm d}
\newcommand\Fi{\rm F_{1}}
\newcommand\Fo{\rm F_{\rm o}}
\newcommand\mcR{\mathcal R}
\newcommand\Prot{\Theta_0}

\begin{document}
\title{Optimal Control of Rotary Motors}
\author{Joseph N. E. Lucero}
\author{Aliakbar Mehdizadeh}
\altaffiliation{Current address: Department of Philosophy of Science, Sharif University of Science and Technology, Tehran, Iran}
\author{David A. Sivak}
\email{dsivak@sfu.ca}
\affiliation{Department of Physics, Simon Fraser University, Burnaby, BC, V5A1S6 Canada}
\date{\today}

\begin{abstract}
Single-molecule experiments have found near-perfect thermodynamic efficiency in the rotary motor $\rm F_1$-ATP synthase.  
To help elucidate the principles underlying nonequilibrium energetic efficiency in such stochastic machines, we investigate driving protocols that minimize dissipation near equilibrium in a simple model rotary mechanochemical motor, as determined by a generalized friction coefficient.
Our simple model has a periodic friction coefficient that peaks near system energy barriers. This implies a minimum-dissipation protocol that proceeds rapidly when the system is overwhelmingly in a single macrostate, but slows significantly near energy barriers, thereby harnessing thermal fluctuations to kick the system over energy barriers with minimal work input. This model also manifests a phenomenon not seen in otherwise similar non-periodic systems: sufficiently fast protocols can effectively lap the system. 
While this leads to a tradeoff between accuracy of driving and energetic cost, we find that our designed protocols out-perform naive protocols.
\end{abstract}

\maketitle

\section{Introduction}
Molecular motors are nanometer-scale protein complexes that convert between different forms of energy. Nature has evolved such motors to play key roles in many essential cellular processes~\cite{Kolomeisky2013}. For example, three families of molecular motors (myosin, kinesin, and dynein) use chemical free energy to drive their translational motion and thereby accomplish useful work, powering a wide variety of intracellular movements~\cite{Schliwa2003}. 

$\Fo\Fi$-ATP synthase~\cite{ATP} is of central importance in virtually all living things, because it synthesizes the ATP molecules that myriad other biological components consume to drive otherwise unfavorable reactions. ATP synthase couples to an electrochemical proton gradient to drive the synthesis of high free energy ATP from low free energy ADP and inorganic phosphate. Despite $\Fo\Fi$'s complex molecular structure, the coupling is mechanically simple~\cite{Coupling} (compared to another rotary biomolecular motor, the bacterial flagellum~\cite{Minamino:2015hd}), with a rotating crankshaft connecting the integral membrane $\Fo$ component to the soluble $\Fi$ and thereby communicating work. 
Various single-molecule experiments monitor $\Fi$'s spontaneous behavior~\cite{Single-Molecule1,Single-Molecule2,Single-Molecule3,Ueno2005} or probe its driven behavior by perturbing $\Fi$ to force rotation of the central crankshaft
For example, with a magnetic bead attached to the central crankshaft, a rotating magnetic field can force rotation of the bead which, in turn, forces rotation of the crankshaft~\cite{Watanabe2011,Saita2015}. 

Among many conceivable constraints, these biomolecular motors face a functional incentive to reduce the unnecessary losses during their energy transduction processes. 
For example, a substantial fraction of energy consumption runs through ATP synthase~\cite{Rolfe:1997vq}, indicating that efficient operation can significantly impact the total cellular energy consumption.  
And indeed, empirically $\Fi$-ATP synthase has been found to be remarkably efficient, approaching 100\%~\cite{ATP-Efficiency}, albeit this mechanical efficiency was experimentally measured in a highly viscous medium near stall torque. 

But ATP synthase must be far from equilibrium to productively function {\em in vivo}, synthesizing more ATP than it hydrolyzes and doing so at a rate sufficient to maintain cellular ATP stores.  
Indeed all biological motors must break detailed balance to achieve directed motion~\cite{feynman66,jarzynski99}.  
So appeals to quasistatic limits may be less relevant for understanding the evolved behavior of these central biological energy transducers.  
Yet we generally lack even a scale on which to understand motor efficiency far from equilibrium.

Moreover, at ambient temperatures these nanoscale protein-based machines experience large conformational fluctuations. Such fluctuations are inherent to their motion, producing experimentally measurable behavior ranging from side steps to reverse operation~\cite{Backward}. 

Thus, there is great interest in characterizing and understanding the fundamental physical limits on the nonequilibrium operation of strongly fluctuating nanoscale systems.  Bioengineers would benefit from systematic understanding of how to harness, rather than oppose, the irresistible influence of thermal fluctuations~\cite{Harness}.

In this paper, we examine molecular motor efficiency by asking: how should a motor be driven to a new conformational ensemble in a limited time, at minimal energetic cost?
This may capture elements of the essential biophysical problem facing $\Fo$, which accrues selective advantage when it can drive $\Fi$ to synthesize ATP at reduced cost, quantified by the protons run down the cross-membrane electrochemical potential difference.
In particular, we study driving protocols that reduce the average work required to rapidly drive a strongly fluctuating system. 
To this end, we make use of a recently developed near-equilibrium theoretical framework that quantifies a system's nonequilibrium response to a time-dependent external perturbation, and hence its dissipation, in terms of a generalized friction coefficient~\cite{SivakOptimalPath,MandalOptimal,Optimal1,Optimal2,Optimal3,Optimal4,Optimal5,Sivak:2016:PhysRevE}. 

We apply this theoretical framework to identify the minimum-dissipation driving protocol for a simple model system of a stochastic rotary molecular machine, that represents the essential physics of recent single-molecule experiments on $\Fi$-ATP synthase~\cite{Mechanically_Driven} and its chemical reaction cycle. 
Previous work~\cite{Sivak:2016:PhysRevE} examined a single barrier crossing; here we extend that study to a series of barrier crossings with periodic boundary conditions.
Similar to the previous work, we find that the designed protocol proceeds rapidly where the system quickly relaxes within a single mechanical basin. 
Conversely, the protocol proceeds slowly when crossing the energetic barrier separating metastable states. 
However, in contrast, the introduction of a periodic potential introduces tradeoffs between work and machine flux not seen in other examined systems.
Despite these tradeoffs, the designed protocols nevertheless often improve upon simple naive driving protocols, particularly at modest driving speeds.

\section{Theory}

The nonequilibrium excess power $\mathcal{P}_{\rm ex}(t_0)$ at time $t_0$ along a driving protocol $\Lambda$ (a dynamic time course of a control parameter $\lambda$) is approximated near equilibrium by~\cite{SivakOptimalPath}
\begin{equation}
\mathcal{P}_{\rm ex}(t_0) \approx \zeta(\lambda(t_0)) \bigg ( \dfrac{\md \lambda}{\md t} \bigg )_{t_0}^{2} \ .
\label{eq:Pex}
\end{equation}
Here the generalized friction coefficient,
\begin{align}
\zeta(\lambda(t_{0})) &= \beta \int_{0}^{ \infty} \md t' \langle \delta f(0) \delta f(t') \rangle_{\lambda(t_{0})}\ ,
\label{eq:form_friction}
\end{align}
is the time integral of the force autocovariance function $\langle \delta f(0) \delta f(t') \rangle_{\lambda(t_{0})}$, defined in terms of the equilibrium fluctuations $\delta f \equiv f - \langle f\rangle_{\lambda(t_{0})}$ of conjugate force $f \equiv - \frac{\partial E(\lambda)}{\partial \lambda}$. $\beta \equiv (k_{\rm B}T)^{-1}$ is the inverse temperature. Angled brackets with subscript $\lambda(t_0)$ denote an ensemble average for equilibrium probability distribution $\bm{\pi}$ at fixed control parameter. \eqref{eq:Pex} is most accurate in the limit of slow driving, when the system remains close to its equilibrium distribution.  

The friction coefficient can be decomposed into
\begin{align}
\zeta(t_{0}) &= \beta \; t_{\rm relax}\left[\lambda(t_{0})\right]\;\langle \delta f^{2}\rangle_{\lambda(t_{0})} \ ,
\label{eq:diff_form_friction}
\end{align}
for equilibrium force variance $\langle \delta f^2 \rangle \equiv \langle \delta f(0)\delta f(0)\rangle_{\lambda(t_{0})}$ at fixed control parameter $\lambda(t_0)$, and integral relaxation time 
\begin{align}
t_{\rm relax}\left[\lambda(t_{0})\right] &\equiv \beta \int_{0}^{ \infty} \md t' \dfrac{\langle \delta f(0) \delta f(t') \rangle_{\lambda(t_{0})}}{\langle \delta f^2 \rangle_{\lambda(t_{0})}} \ .\label{eq:relax_time}
\end{align}
A \emph{designed} protocol (chosen to minimize dissipation within this linear-response framework) proceeds at a velocity proportional to the inverse square root of the generalized friction coefficient $\zeta(\lambda(t))$~\cite{SivakOptimalPath}, 
\begin{equation}
\dfrac{\md }{\md t}\lambda^{\rm designed}(t) \propto \zeta(\lambda(t))^{-(1/2)} \ ,
\label{eq:designed_vel}
\end{equation}
where the proportionality is set by requiring the protocol complete in duration $t_{\Lambda}$. 
Substituting~\eqref{eq:designed_vel} into \eqref{eq:Pex} results in excess power being constant across a designed protocol. 
Note that exact solutions for the minimum-work protocol (not making the linear-response approximation) are only known in extremely simple systems, harmonic potentials~\cite{Optimal3,Optimal4}.

\section{Model}
In this work we take inspiration from $\Fi$-ATP synthase; however, these results in principle generalize to any other periodic motor that couples mechanics and chemistry (e.g., kinesin~\cite{Valentine2006}). 
Experimentally, $\Fi$-ATP synthase can be driven by a time-dependent sinusoidal potential due to the interaction between a magnetic trap and a magnetic bead attached to the motor's central crankshaft~\cite{Mechanically_Driven}. The magnetic trap can be rotated to drive the magnetic bead, and hence the attached crankshaft, over a sequence of rotational mechanical energy barriers separating the intrinsic metastable states of the motor. 

\begin{figure*}[!ht]
	\subfloat{
		\hspace*{-1.1cm}\includegraphics[clip,width=1.1\columnwidth]{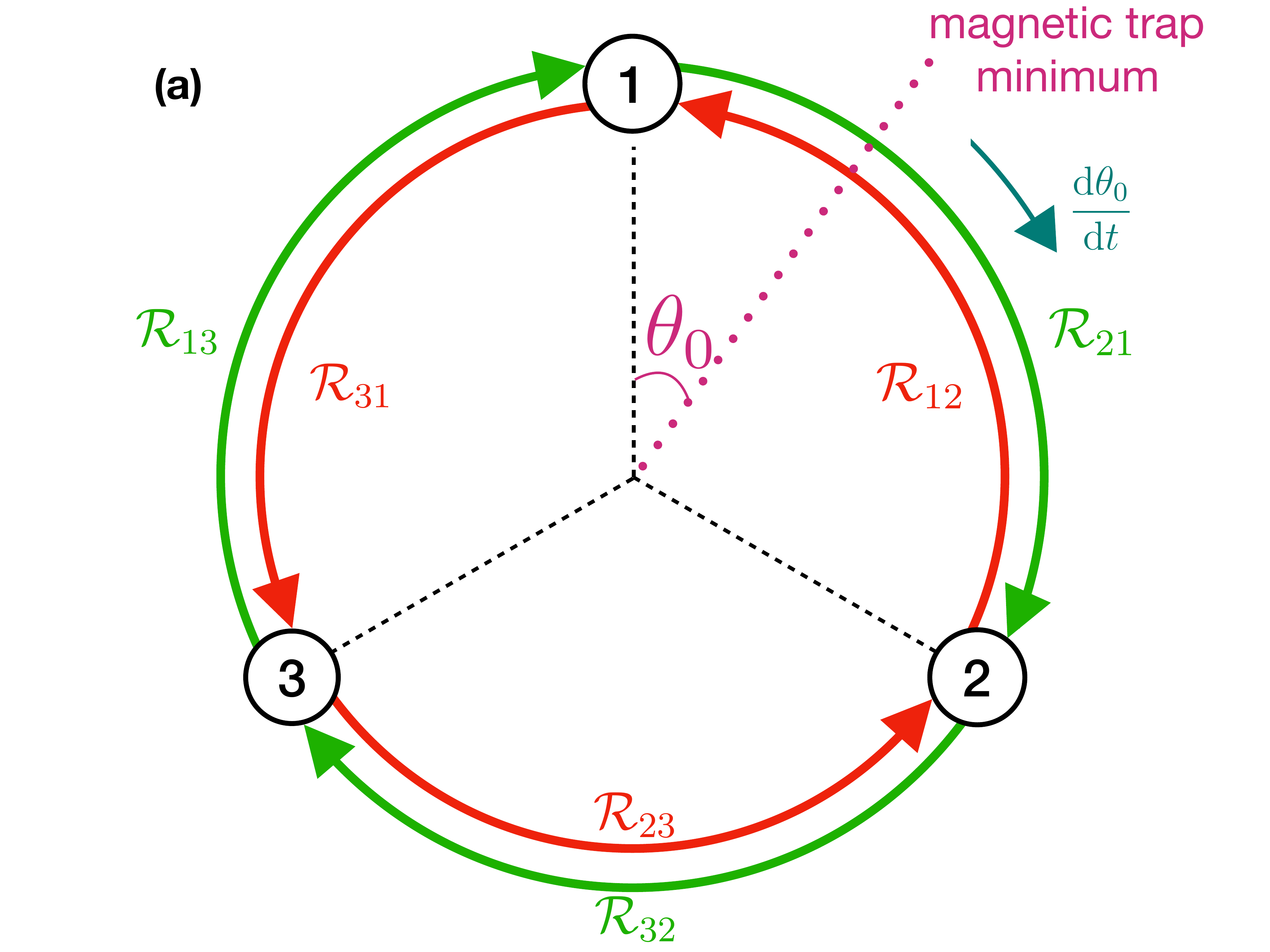}
		\label{fig:Model}
	}
	\subfloat{
		\hspace*{0.1cm}\includegraphics[clip,width=1.045\columnwidth]{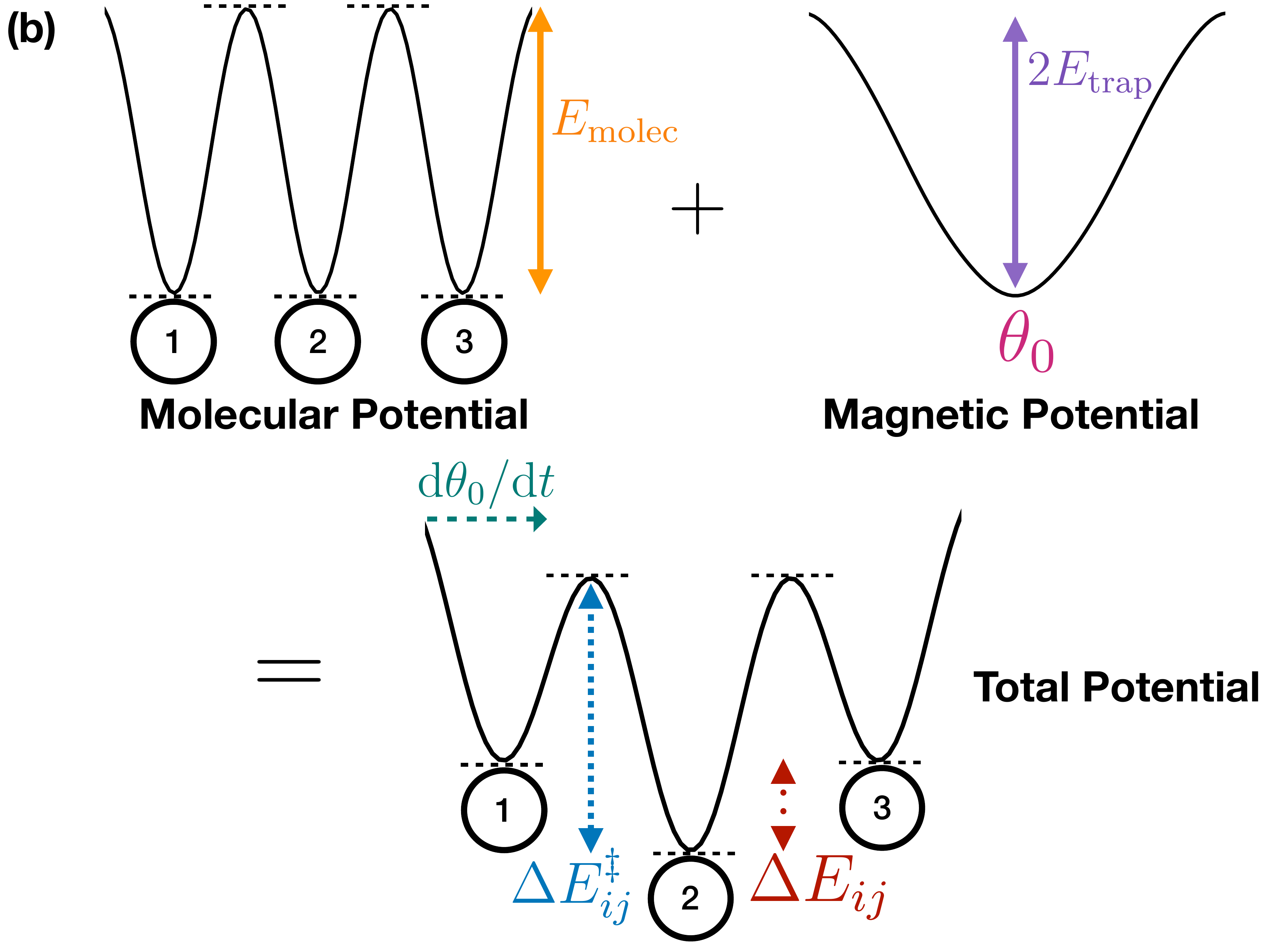}
		\label{fig:Energy_Landscape}		
	}
	\caption{(a) A simple model of a rotary mechanochemical machine driven by a magnetic trap. 
	The conformational states $i=1$, $i=2$, $i=3$ (circles) have angular locations $ \theta = 0 , \ 2\pi/3, \ \text{and}\ 4\pi/3$, respectively. 
	Transitions correspond to crankshaft rotation. 
	The magnetic trap is parameterized by the angular orientation of the trap minimum $\theta_{0}$ (the control parameter) relative to state $i=1$ at $\theta = 0$. 
	(b) Schematic of the intrinsic molecular potential, the magnetic potential, and their superposition that comprises the system's energy landscape. 
	Mechanical barriers of height $\Delta E^{\ddagger}_{ij}(t)$ \eqref{eq:mech_difference} at $\pi/3$, $\pi$, and $5\pi/3$ (square-dotted blue arrow) and energy differences $\Delta E_{ij}(t)$~\eqref{eq:state_difference} of the conformational states (circular-dotted red arrow) modulate transition rates between states $i$ and $i+1$ at time $t$. 
	}
	\label{fig:Land_Model}
\end{figure*}

At moderate temporal resolution, current experiments find three metastable crankshaft angles separated by $2\pi/3$~\cite{ATP-Efficiency}. Therefore, we model the molecular potential as consisting of three states on a ring (Fig.~\ref{fig:Model}). Our coarse-grained model only resolves the mechanical rotational steps, corresponding to $i\in\{1,2,3\}$. We implicitly assume a tight coupling between the mechanical motion of $\Fi$ and the chemical steps of synthesis/hydrolysis and binding/unbinding of ATP and its hydrolysis products. We also implicitly hold chemical concentrations fixed at equilibrium ratios. 

Thus, we simulate the state of our system as a particle diffusing over a periodic energy landscape comprised of two components: a time-independent sinusoidal molecular potential with periodicity $2\pi/3$ and amplitude $\beta E_{\rm molec}$ (motivated by the statistical mechanics of a system with three metastable conformational mesostates), and a time-dependent sinusoidal magnetic trap potential with periodicity $2\pi$, amplitude $\beta E_{\rm trap}$, and minimum at our control parameter $\theta_{0}$ (Fig.~\ref{fig:Energy_Landscape}).

The external time-dependent driving force, a rotating magnetic trap, is modeled by a sinusoidal potential. The magnetic trap imposes a preferred orientation $\theta_0$ (corresponding to the minimum of the periodic potential) on the magnetic bead, that is changed according to a deterministic rotation schedule specified by \eqref{eq:designed_vel}. The rotation schedule defines a given protocol $\Prot$.
(For simplicity, we restrict our analysis here to the optimization of a single control parameter---the rotational angle of a magnetic trap---but one could generalize this study to simultaneous variation of multiple control parameters, such as magnetic trap angle and stiffness, using multidimensional optimization methods applied to more complex systems~\cite{Ising}.) 

A three-dimensional probability vector $\mathbf{p}(t)$ represents the probability distribution over system microstates at time $t$. Variation of the control parameter drives the system distribution away from equilibrium, with a time-dependent evolution according to the master equation~\cite{Aldous2002,vanKampen,Jacobs2010},
\begin{align}
\dfrac{\md \mathbf{p}(t)}{\md t} =  \mathbf{p}(t)\bm{\mcR}(t),
\label{eq:master_equation}
\end{align}
where $\bm{\mcR}(t)$ is a tridiagonal $3 \times 3$ transition rate matrix representing the system's driven diffusive dynamics. For $i \neq j$, the matrix element $\mcR_{ij}(t)$ is the instantaneous rate at which probability flows from state $ i $ to state $ j $ at time $ t $. Conservation of probability determines the diagonal elements $\mcR_{ii}(t) = -\sum_{j\neq i}\mcR_{ij}(t)$. 

Due to the three-fold symmetry of $\Fi$, we use equal unperturbed energies for each of the system's three microstates. Choosing this unperturbed energy as the zero of the system's energy, the transition rate matrix elements, as well as the system's energy profile, depend on the external sinusoidal potential according to 
\begin{subequations}
	\begin{align}
	\mcR_{i,i\pm1}(t) &= \min\left(1, e^{-\beta \Delta E^{\ddag}_{i,i\pm1}(t)}, e^{-\beta \Delta E_{i,i\pm1}(t)}\right) \label{eq:transition_rates} \\
	\Delta E_{i,i\pm1}^{\ddag}(t) &= E_{\rm molec} \label{eq:mech_difference}\\
    &- E_{\rm trap}\left[\cos\left(\theta_{i\pm 1}^{\ddag}-\theta_{0}(t)\right)-\cos\left(\theta_{i} - \theta_{0}(t)\right)\right]\nonumber\\
	\Delta E_{i,i\pm 1}(t) &= -E_{\rm trap}\left[\cos\left(\theta_{i\pm1}-\theta_{0}(t)\right) - \cos\left(\theta_{i} - \theta_{0}(t)\right)\right] \ .\label{eq:state_difference}
	\end{align}
	\label{eq:sum-rate}
\end{subequations} 
To simplify this model system and reduce the parameter space, we enforce the additional symmetry that mechanical energy barriers are halfway between their corresponding metastable states, at $\theta_{i,(i+1)\!\!\!\mod 3}^{\ddag} = (2i-1)\frac{\pi}{3}$.

The system evolution is simulated over one full rotation of the control parameter $\theta_0$ (ie. $\theta_{0}(t) \in [0,2\pi]$) lasting a duration $t_{\Prot}$. The system begins at time $t_{0} = 0$ in periodic steady state, and the control parameter $\theta_{0}$ evolves according to the specified protocol. Appendix~\ref{sec:app_sim_details} gives more details of the numeric calculations.

We explore a parameter regime that includes a slowly varying magnetic trap manipulation of a three-state machine with similar coarse-grained kinetics to $\rm F_1$. Here, the rotational diffusion coefficient is dominated by the contribution from the micron-sized magnetic bead, since it is considerably larger than the crankshaft. A typical magnetic bead has radius $r = 275\; \rm nm$~\cite{You2012}, and is tethered via a streptavidin-biotin link to the machine's outer surface. The modified Stokes hydrodynamic rotational diffusion coefficient for such a bead rotationally diffusing around its edge is $D_{\rm bead} \approx 4.5\;\si{\square\radian\per\second}$~\cite{Xu2008}. By contrast, the hydrodynamic diffusion coefficient of the isolated $\rm F_1$'s central shaft has been estimated as $D_{\rm shaft} \approx \num{7e7}\;\si{\square\radian\per\second}$~\cite{DiffusionEst}. Appendix~\ref{sec:app_time_scale} gives details on dynamical time scales.

\section{Results}

\subsection{Friction coefficient}

\begin{figure}[!ht]
	\subfloat{
		\includegraphics[clip,width=0.96\columnwidth]{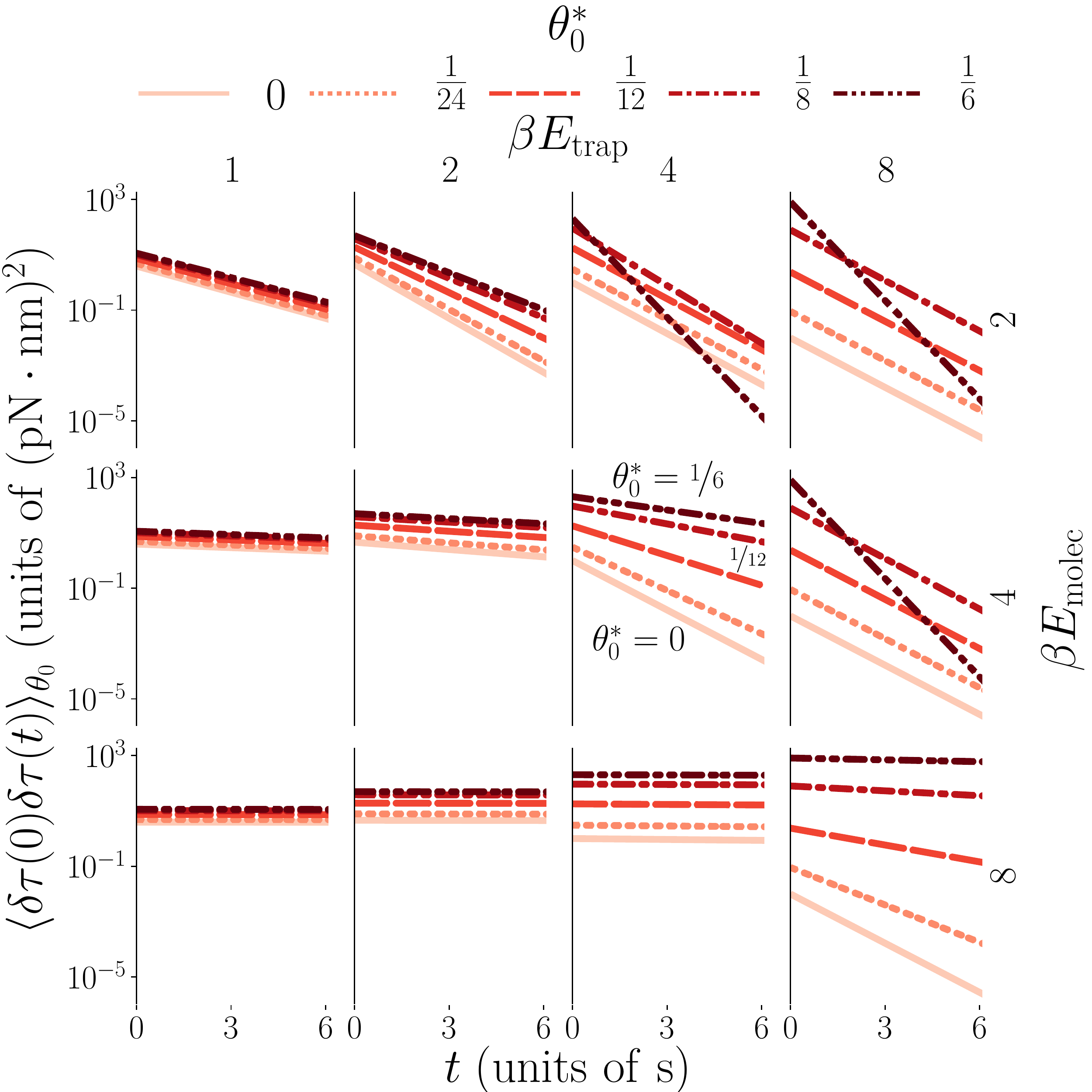}
		\label{fig:autocovar}
	} 
	\caption{
	Equilibrium autocovariance $\langle \delta\tau(0) \delta\tau(t) \rangle_{\theta_{0}}$ of torque fluctuations $\delta\tau$ exerted on the magnetic bead by external magnetic fields, across a range of times (within a plot), for different values of the dimensionless control parameter $\theta_{0}^{*} \equiv \theta_{0}/2\pi$ (different curves within a given subplot), varying perturbation strength $\beta E_{\rm trap}$ across columns, and varying intrinsic barrier height $\beta E_{\rm molec}$ across rows. 
	Appendix~\ref{sec:autocorr_calculate} shows that all autocovariance curves do eventually decay to negligible levels, though on quite disparate timescales. 
    Appendix~\ref{sec:app_symmetry} discusses the anomalously fast relaxation timescale at some magnetic field orientations, for a strong trap relative to barrier height.
    }
	\label{fig:TheAutocovar}
\end{figure}

For our model, with the control parameter being the energetic minimum of the magnetic trap, the conjugate force $f$ is the torque $\tau$ arising from the external magnetic fields interacting with the magnetic bead. Figure~\ref{fig:TheAutocovar} shows the torque autocovariance $\langle \delta\tau(0) \delta\tau(t) \rangle_{\theta_{0}}$ as a function of fixed 
(dimensionless) control parameter $\theta_{0}^{*} \equiv \theta_{0}/2\pi$, for given magnetic trap strength $\beta E_{\rm trap}$ and intrinsic mechanical barrier $\beta E_{\rm molec}$.    

\begin{figure}[!ht]
	\subfloat{
		\includegraphics[clip,width=0.96\columnwidth]{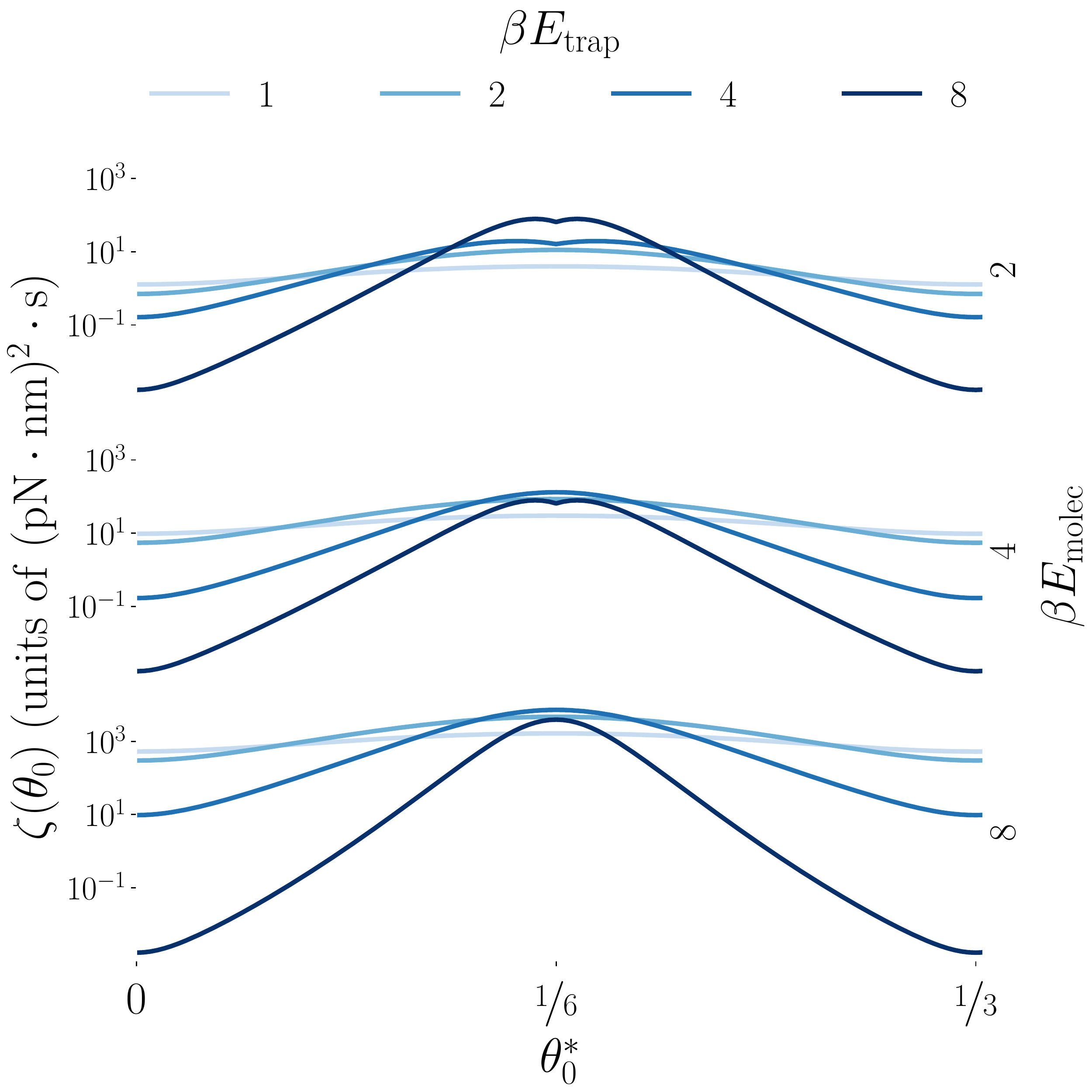} 
		\label{fig:friction}
	}
	\caption{
	Generalized friction coefficient $\zeta(\theta_{0})$ across a range of trap minima $\theta_{0}^{*} \equiv \theta_{0}/2\pi$ (within a curve), for varying perturbation strength $\beta E_{\rm trap}$ (different shades of curves in a given subplot), and varying intrinsic barrier height $\beta E_{\rm molec}$ across rows. 
	Appendix~\ref{sec:app_symmetry} discusses the dip in the friction coefficient at the barrier, for strong trap strength relative to barrier height.
	}
	\label{fig:TheGFriction}
\end{figure}

Integrating the torque autocovariance function~\eqref{eq:diff_form_friction} yields the generalized friction coefficient. The structural threefold symmetry of our model produces a friction coefficient $\zeta(\theta_{0})$ that is a periodic function of $\theta_{0}$ with a periodicity of $2\pi/3$ (Fig.~\ref{fig:TheGFriction}). We thus focus on the first $2\pi/3$ period since the other periodic images give identical results. 

Figure~\ref{fig:TheGFriction} shows that the generalized friction coefficient varies significantly with varying trap strength $\beta E_{\rm trap}$, mechanical barrier $\beta E_{\rm molec}$, and trap minimum $\theta_{0}^{*}$. 
For low barrier heights and weak magnetic traps, the friction coefficient is relatively small and flat, not varying with the trap's minimum. 
For increasing trap strength, the friction coefficient increasingly varies with trap minimum. 
In particular, when a strong magnetic trap is applied, across a given landscape the generalized friction coefficient can vary significantly, by more than two orders of magnitude.

Intuitively, a large friction coefficient is produced either when two metastable states are of comparable energy (so equally probable) leading to a large torque variance, and/or for large integral relaxation time. For our system, both these conditions generally occur when the trap minimum is at the barrier location, $\theta_{0} = \theta_{i(i+1)}^{\ddag}$ (except when $E_{\rm trap} \ge 2E_{\rm molec}$, due to our system's coarse-graining [see Appendix~\ref{sec:app_symmetry}]).

\subsection{Minimum-dissipation driving velocity and protocol}

\begin{figure}[!ht]
	\subfloat{
		\includegraphics[clip,width=0.96\columnwidth]{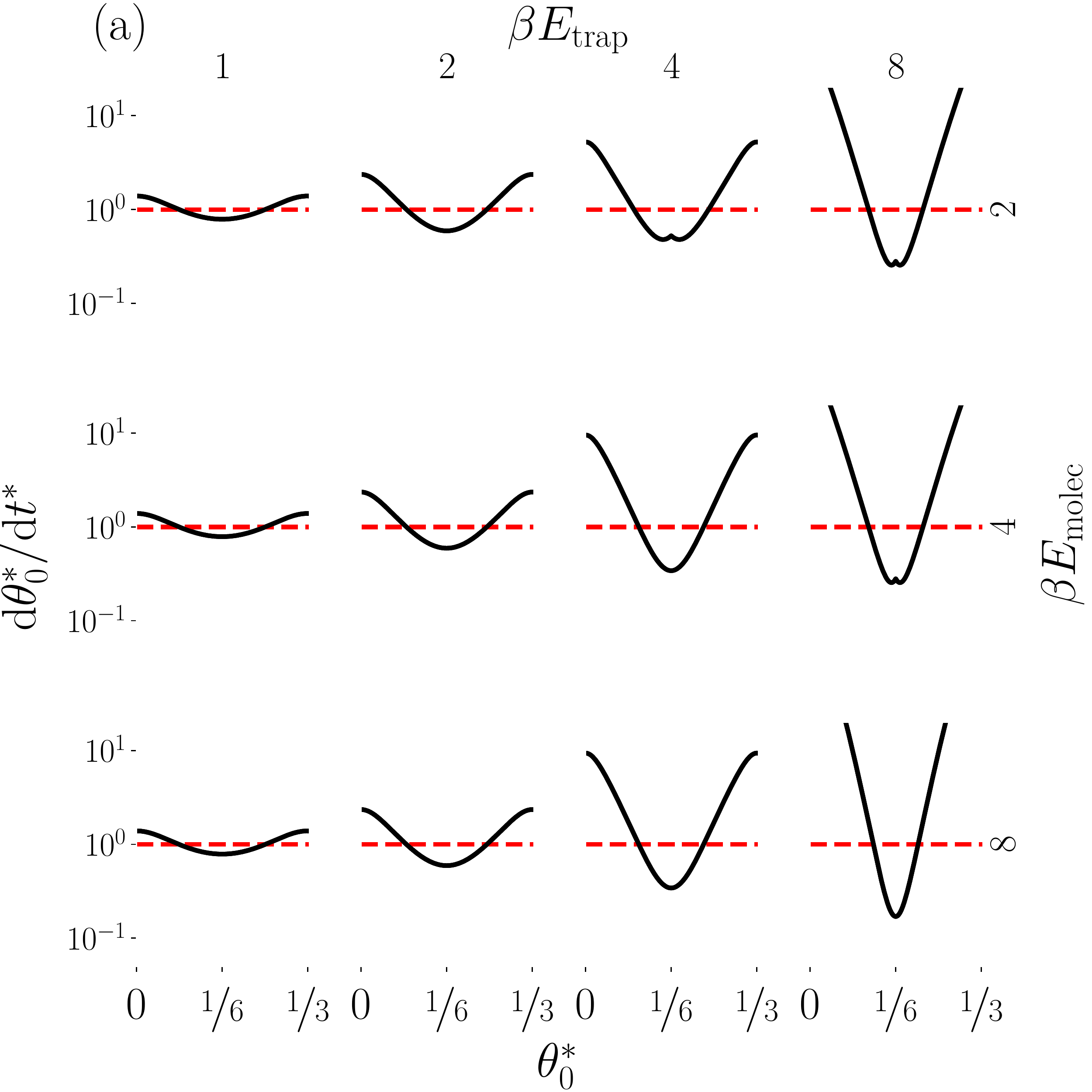}
		\label{fig:normed-vel}
	}
	
	\subfloat{
		\includegraphics[clip,width=0.96\columnwidth]{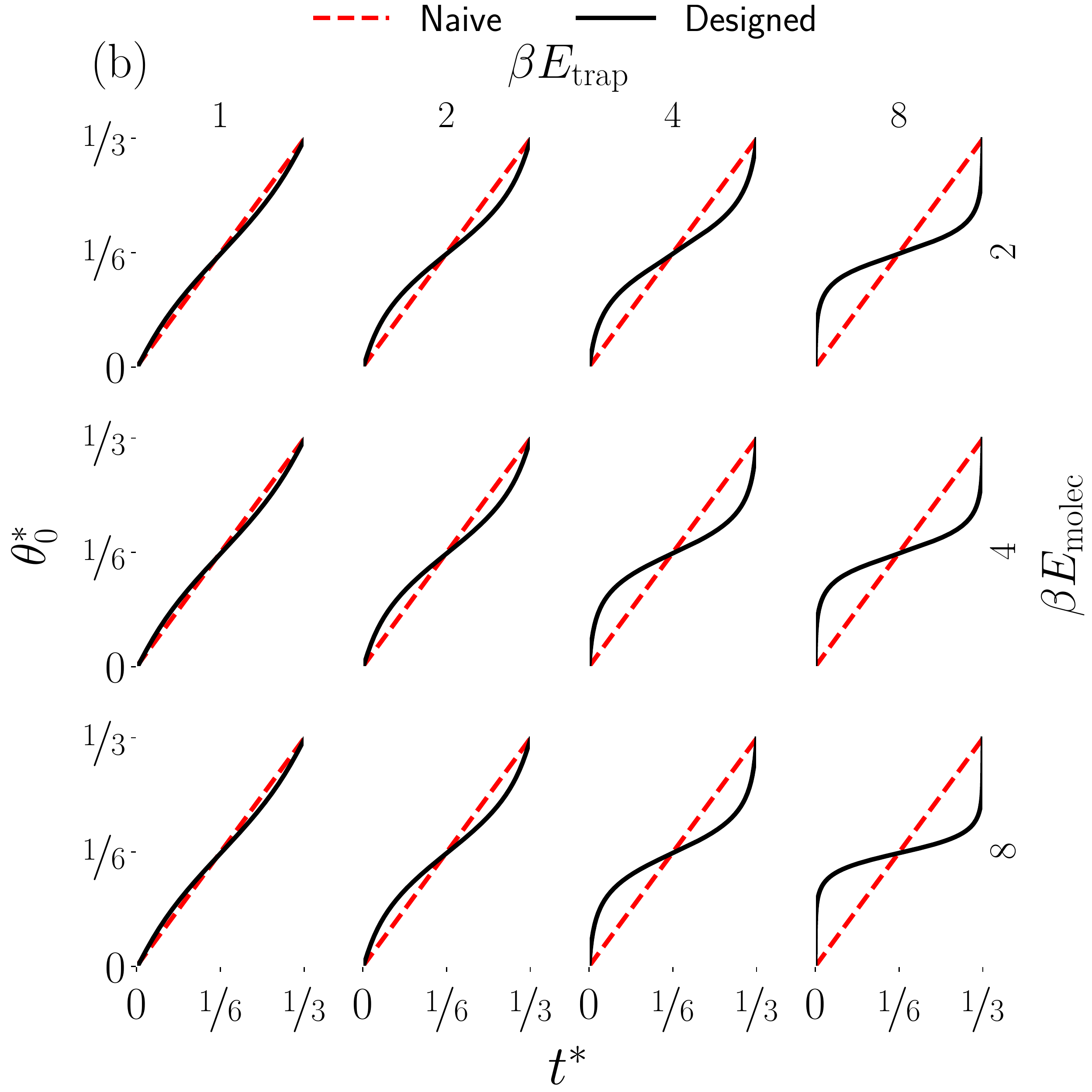} 
		\label{fig:normed-path}
	}
	\caption{(a) (Dimensionless) Control parameter velocity $\md \theta_{0}^{*} / \md t^{*}$, as a function of control parameter $\theta_{0}^{*} \equiv \theta_{0}/2\pi$, on a naive protocol (red dashed lines) and designed protocol (black solid lines). 
		(b) Control parameter $\theta_{0}^{*}$ as a function of dimensionless time $t^{*} \equiv t / t_{\Prot}$ for naive protocols (red dashed lines) and designed protocols (black solid lines). Same variation of $\beta E_{\rm trap}$ and $\beta E_{\rm molec}$ as in Fig.~\ref{fig:TheAutocovar}.}
	\label{fig:vel-path}	
\end{figure}

Given the variation of the generalized friction coefficient over as much as several orders of magnitude, the designed driving velocity~\eqref{eq:designed_vel} also varies by orders of magnitude across a given designed protocol, especially when employing a strong magnetic trap (Fig.~\ref{fig:normed-vel}). 

The driving velocity profile across control parameter space is invariant under changes in protocol duration $t_{\Prot}$, up to a scaling constant to ensure that the protocol, consisting of a full $2\pi$ rotation of the control parameter, completes in the allotted duration. Therefore, here we present our results with respect to the dimensionless time $t^{*} \equiv t/t_{\Prot}$.

As energy barriers increase (top to bottom rows in Fig.~\ref{fig:normed-vel}), the designed driving velocity near the energy barrier decreases, indicating that near high energy barriers, the system should be driven slow in order to keep the instantaneous nonequilibrium probability distribution more similar to the equilibrium distribution. 
Physically, this slowdown near the barrier arises from the fact that, in order to minimize dissipation, the designed protocol is constructed such that it allocates most of its time near the barrier so as to let thermal fluctuations aid the system in jumping over the energy barrier.

Integrating the velocity gives the control parameter as a function of time (Fig.~\ref{fig:normed-path}). For low energy barriers relative to trap strength ($E_{\rm trap} \ll E_{\rm molec}$), when the friction coefficient is relatively flat, the designed protocol doesn't differ significantly from the naive protocol, so no significant energy is saved by performing the designed protocol; however, when the friction coefficient varies strongly across control parameter space, as for $E_{\rm trap} \gg E_{\rm molec}$, the designed protocol deviates significantly from the naive protocol.

\subsection{Excess power} 

\begin{figure}[!ht]
	\subfloat
	{
		\includegraphics[clip,width=0.96\columnwidth]{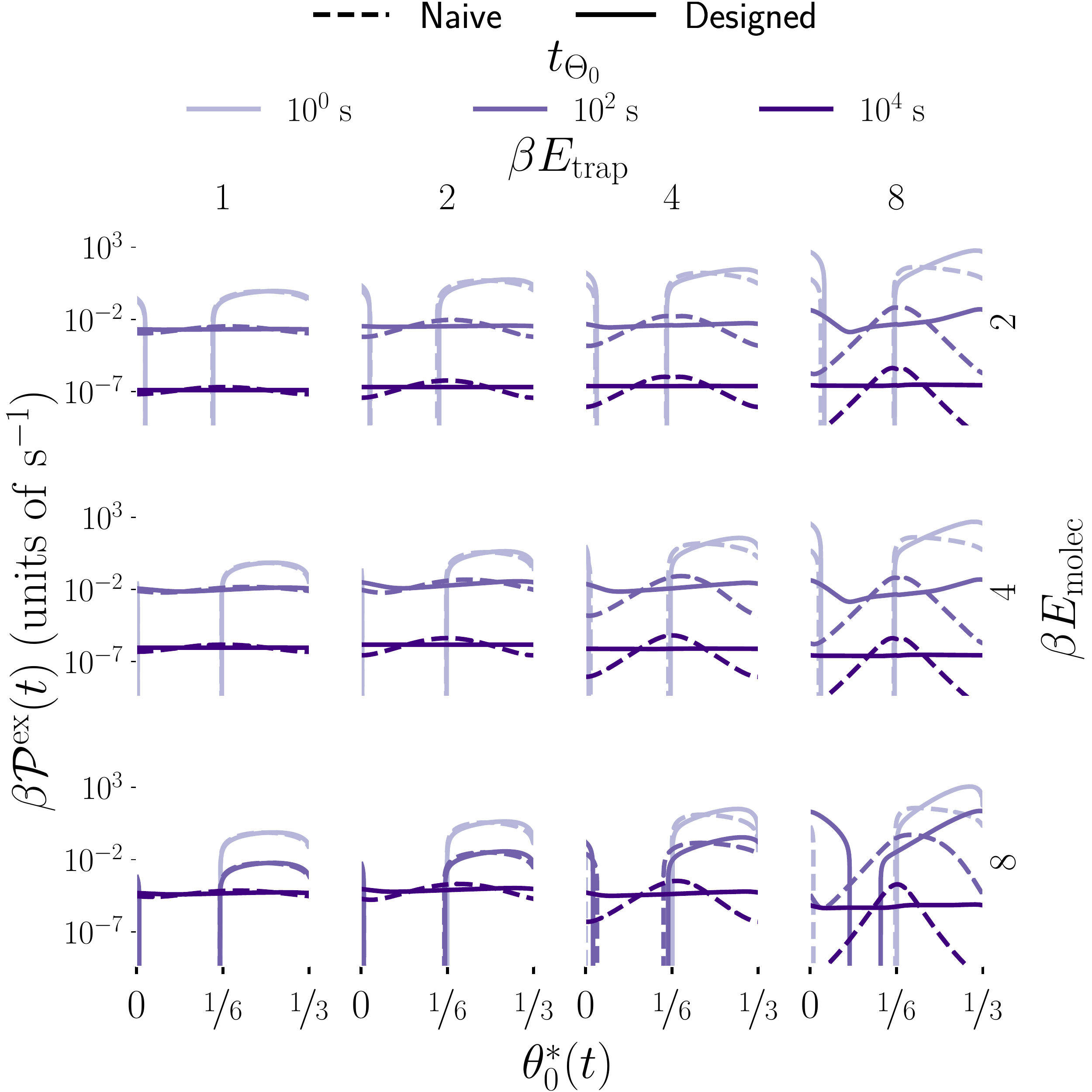} 
	}
	\caption{
	Instantaneous excess power $\beta \mathcal{P}^{\rm ex}(t)$, on naive protocols (dashed lines) and designed protocols (solid lines), for a variety of protocol durations $ t_{\Prot} $ (different shades), as a function of instantaneous trap minimum $\theta_{0}^{*}(t)$. 
	Excess power is calculated exactly via numerical master equation solutions that do not assume linear response. Same variation of $\beta E_{\rm trap}$ and $\beta E_{\rm molec}$ as in Fig.~\ref{fig:TheAutocovar}
	}
	\label{fig:naive_optimal_powers}
\end{figure}

Figure~\ref{fig:naive_optimal_powers} shows the excess power as a function of trap minimum.  Note that---as in Fig.~\ref{fig:TheGFriction}---we show data from $\theta_0^* \in [0,\tfrac{1}{3}]$, because $\theta_0^* \in [\tfrac{1}{3},\tfrac{2}{3}]$ and $\theta_0^* \in [\tfrac{2}{3},1]$ give identical results.  

During naive protocols with sufficiently long protocol duration, the excess power varies by multiple orders of magnitude and is periodic (darkest dashed curves). 
Conversely, designed protocols---still with long protocol durations---have constant excess power across the protocol (darkest solid curves in Fig.~\ref{fig:naive_optimal_powers}).  

In general, the linear-response approximation~\eqref{eq:Pex} works well when the system remains close to equilibrium throughout the entire protocol. For driven systems this is satisfied when the protocol duration is long (relative to the relaxation timescale of the system) and hence when the driving velocity is slow. Appendix~\ref{sec:approx_to_exact} compares in detail the linear-response approximation and the exact system behavior. 

If the protocol proceeds sufficiently slowly that the system remains close to equilibrium throughout the protocol (darkest curves in Fig.~\ref{fig:naive_optimal_powers}), the excess power near a barrier closely matches that of a bistable system (a two-state system separated by an energetic barrier) driven by a harmonic trap with time-dependent minimum~\cite{Sivak:2016:PhysRevE}. In this case, the distant high-energy microstate and landscape periodicity have negligible effect on the current distribution over microstates. When the system is driven further from equilibrium, the nonlinear time-dependent external force and periodic landscape lead to departures from the bistable system behavior. 

In both the naive and designed protocols, 
for sufficiently short protocol duration (sufficiently high velocity), and particularly for $E_{\rm trap} \ll E_{\rm molec}$, the excess power can become negative (curve disappears on a logarithmic plot).
This results from the periodic boundary conditions, as the control parameter `catches up' to the system.
If the protocol is sufficiently short and the trap is sufficiently weak, then in the first half of the protocol ($\theta_{0}^{*} \in \left[0,\tfrac{1}{2}\right)$), a large portion of the initial system probability cannot cross the first barrier and is therefore left behind in the starting basin, thus the trap does excess work on the system. In the second half of the protocol ($\theta_{0}^{*} \in \left[\tfrac{1}{2},1\right]$), the trap moves towards the starting state. With much of the system probability still in the starting state, as the trap nears a complete rotation, it `catches up' with the probability density it left behind, so the system does work on the trap, resulting in the observed negative excess power. 

\subsection{Excess work}
\begin{figure}[!ht]
	\subfloat
	{
		\includegraphics[clip,width=0.96\columnwidth]{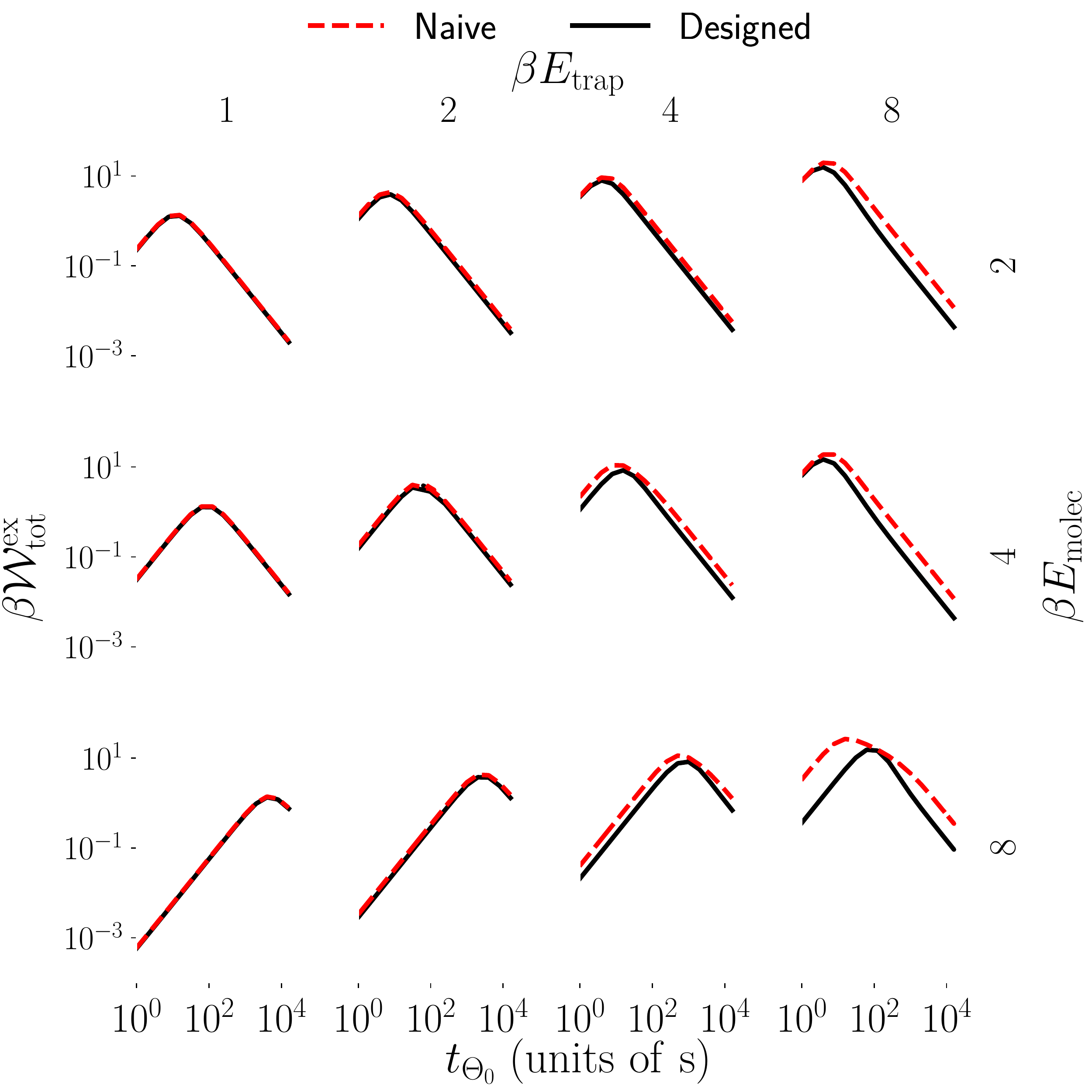} 
	} 
	\caption{
	Total excess work $\beta \mathcal{W}^{\rm ex}_{\rm tot}$ during naive protocols (red dashed lines) and designed protocols (solid black lines), as a function of protocol duration $t_{\Prot}$. 
	Total excess work is calculated exactly from the numerical master equation solutions, not assuming linear-response. Same variation of $\beta E_{\rm trap}$ and $\beta E_{\rm molec}$ as in Fig.~\ref{fig:TheAutocovar}
	}
	\label{fig:naive_optimal_works}
\end{figure}

Figure~\ref{fig:naive_optimal_works} shows that---in the parameter regime we explored---designed protocols require less excess work than naive protocols do.
Moreover, for a given trap strength $\beta E_{\rm trap}$ and molecular barrier height $\beta E_{\rm molec}$, total excess work initially increases linearly as protocol duration increases; however, with further increases of protocol duration, the total excess work decreases linearly. 
The duration $t_{\rm crossover}$---where the excess work is maximized---decreases with increasing trap strength and increases with increasing molecular barrier height. 

Similar to the phenomenon of the negative excess power, this maximum originates from the system's periodic boundary conditions. For $t_{\Prot} < t_{\rm crossover}$, as $t_{\Prot}$ increases, the system has progressively more time to relax in response to the imposed protocol, so more probability follows the protocol, so the protocol does more work on the system. Conversely, for $t_{\Prot} > t_{\rm crossover}$, the system follows the protocol and remains near equilibrium, linear-response provides a good approximation, and so excess work decreases as $1/t_{\Prot}$~\cite{SekimotoSasa1997,Crooks2007}. 
This competition leads to the maximum at $t_{\rm crossover}$.

\subsection{Machine flux}
\label{sec:machine_flux}
\begin{figure}[!ht]
	\subfloat
	{
		\includegraphics[clip,width=0.96\columnwidth]{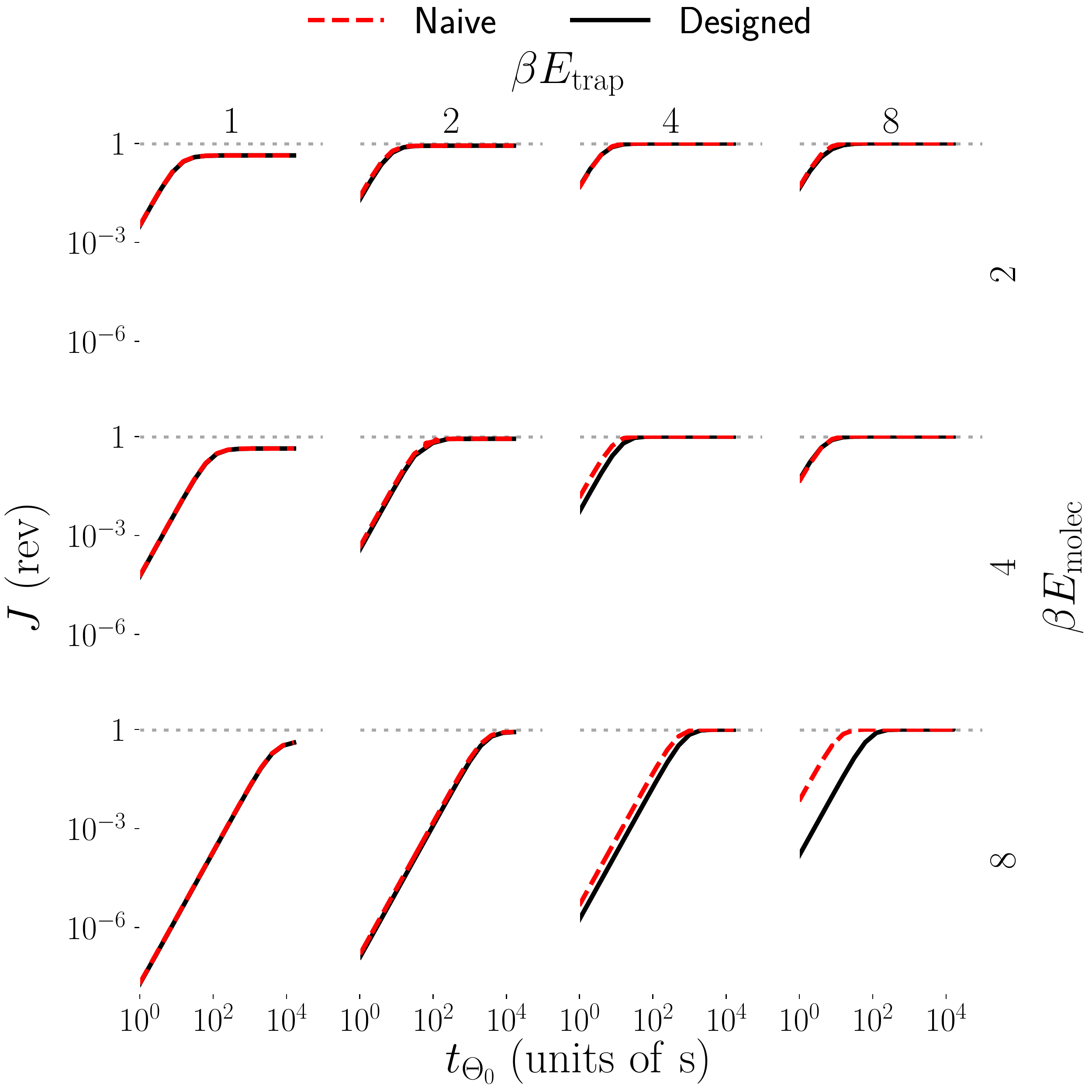} 
	} 
	\caption{Flux $J$ during a complete naive (red dashed lines) or designed protocol (solid black lines), as a function of protocol duration $t_{\Prot}$. Dotted gray line denotes unity.
		Same variation of $\beta E_{\rm trap}$ and $\beta E_{\rm molec}$ as in Fig.~\ref{fig:TheAutocovar}.}
	\label{fig:naive_optimal_fluxes}
\end{figure}

We quantify machine progress by the total average net flux $J$ (hereafter simply \emph{flux}) during the entire protocol covering a complete cycle~\cite{Brown2017,Anandakrishnan2016}. 
Figure~\ref{fig:naive_optimal_fluxes} shows that $J$ approaches zero in the limit of short protocol durations, whereas it achieves a limiting value at sufficiently long protocol durations. This limiting value is unity for large $\beta E_{\rm trap}$.
The duration $t_{\rm limit}$ at which $J$ nears its limiting value decreases with increasing $\beta E_{\rm trap}$ and increases with increasing $\beta E_{\rm molec}$.
The flux only differs substantially between protocol types at the largest $\beta E_{\rm molec}$ and $\beta E_{\rm trap}$ (bottom right subplot).

When the control parameter is at a metastable state, energy barriers are relatively low (compared to adjacent states), thus maximizing flow into the single ground state. 
Since the system lags the control parameter, there is more probability in the previous state than in the next state, so the net flux is forward.  
By contrast, when the control parameter is at the barrier, the barriers are relatively high, limiting flux in either direction.  
Thus instantaneous flux is higher when the control parameter is at a metastable state and lower when at an energetic barrier. 

For short protocol durations, the naive protocol produces a higher flux than the designed protocol for a similar protocol duration.
The designed protocol spends a comparatively small proportion of the protocol duration with the control parameter near a metastable state, so as to increase the time spent near the barrier and thus increase the opportunity for thermal fluctuations to carry the system over the barrier; however, instantaneous flux is maximized when the control parameter is at a basin and thus, for short protocol durations, the designed protocol misses regions of greatest instantaneous flux, producing a lower total flux over the entire protocol. 

\subsection{Distance from equilibrium}
We use the relative entropy (Kullback-Leibler divergence)~\cite{Cover:2006:Book} between nonequilibrium and equilibrium distributions,
\begin{equation}
D\left[\mathbf{p}(t)||\bm{\pi}(\theta_{0}^*(t))\right] \equiv \sum\limits_{i}p_{i}(t)\ln\left[\dfrac{p_{i}(t)}{\pi(\theta_{i}|\theta_{0}^*(t))}\right]\ ,\label{eq:rel_entropy}
\end{equation}
as a physically meaningful measure of difference~\cite{SivakCrooks2012,Gaveau1997},  from the equilibrium distribution 
\begin{align}
\bm{\pi}(\theta_{0}^*(t_{n})) \equiv \left[\ \pi(\theta_{1}|\theta_{0}^*(t_{n})),\  \pi(\theta_{2}|\theta_{0}^*(t_{n})),\ \pi(\theta_{3}|\theta_{0}^*(t_{n}))\ \right]\ ,
\end{align}
because it is proportional to the work available as a system relaxes from the nonequilibrium distribution to the equilibrium distribution. 

\begin{figure}[!ht]
	\subfloat{
		\includegraphics[clip,width=0.96\columnwidth]{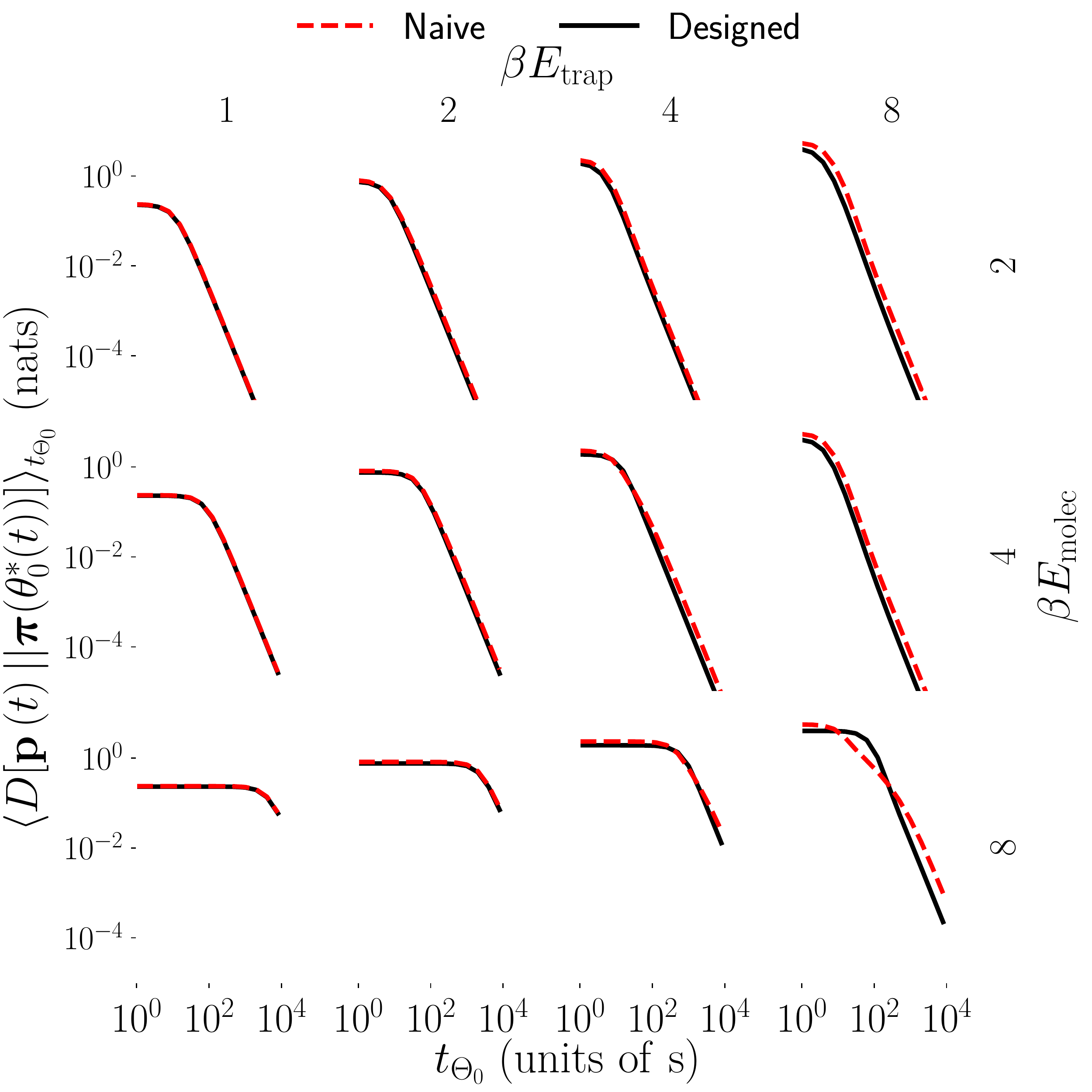} 
	}
	\caption{
	Time-averaged relative entropy (Kullback-Leibler divergence) 
	$\langle D\left[\mathbf{p}(t)||\bm{\pi}(\theta_{0}^*(t))\right]\rangle_{t_{\Prot}}$ between the instantaneous nonequilibrium distribution $\mathbf{p}(t)$ and the equilibrium distribution $\bm{\pi}(\theta_{0}^*(t))$, as a function of protocol duration $t_{\Prot}$, for naive (dashed lines) and designed protocols (solid lines). 
	Same variation of $\beta E_{\rm trap}$ and $\beta E_{\rm molec}$ as in Fig.~\ref{fig:TheAutocovar}.}
	\label{fig:the_mean_distance}
\end{figure}

Intuitively, longer-duration protocols keep the system closer to equilibrium, and thus lead to a smaller relative entropy. Figure~\ref{fig:the_mean_distance} indeed shows that as protocol duration increases, the time-averaged distance from equilibrium $\langle D\left[\mathbf{p}(t)||\bm{\pi}(\theta_{0}^*(t))\right]\rangle_{t_{\Prot}}$ decreases.

Figure~\ref{fig:the_mean_distance} also shows that a designed protocol typically keeps the system closer to equilibrium than the corresponding naive protocol does, with an exception for the strongest trap and highest barrier energy during protocols of intermediate duration. Appendix~\ref{sec:inst_rel_entropy} provides detailed analysis of the instantaneous distance from equilibrium.

For long protocol durations, the designed protocol is on average closer to equilibrium than the naive protocol is. In this regime, the system obeys linear-response, having ample time to respond to the imposed perturbation. This is consistent with the derivation producing \eqref{eq:designed_vel}, predicting that the designed protocol should keep the system closer to equilibrium in the linear-response regime.

For intermediate durations, the designed protocol can drive the system further from equilibrium than the corresponding naive protocol. Here, the system responds significantly to both protocols, as reflected by the flux at these protocol durations being near unity. 
The designed protocol is specifically crafted so that it spends, proportionally, the smallest amount of time at the metastable basins of the system, so as to maximize the amount of time that it spends at the energetic barriers. This increases the likelihood that thermal fluctuations can kick the system over the barrier, and hence reduces the amount of external work required.
However, as discussed in \S~\ref{sec:machine_flux}, this means that the designed protocol spends less time where flux is highest---the metastable basins---and instead spends the most amount of time where the flux is the smallest---the energetic barriers. Avoiding areas where flux is greatest, it can drive the system further from equilibrium. 
Conversely, the naive protocol spends equal proportions of time at the basin and at the barrier, so can keep the system closer to equilibrium. 

For the shortest protocol durations, the designed protocol is on average closer to equilibrium than the naive protocol is. As seen in Fig.~\ref{fig:naive_optimal_fluxes}, at high $\beta E_{\rm trap}$ and $\beta E_{\rm molec}$ the naive protocol produces greater flux---as much as two orders of magnitude greater---than the corresponding designed protocol. 
For this rapid driving, the extra flux in the naive protocol takes the form of the system following---but persistently lagging---the control parameter, thus increasing the distance from equilibrium compared to the designed protocol where the system is relatively unresponsive, hence in a periodic landscape equally likely to be lagging or `ahead of' the control parameter.  
For low $\beta E_{\rm trap}$ or $\beta E_{\rm molec}$, the two protocol types produce comparable flux and comparable distance from equilibrium.

\subsection{Coefficient of performance}

\begin{figure}[!ht]
	\subfloat{
		\includegraphics[clip,width=0.96\columnwidth]{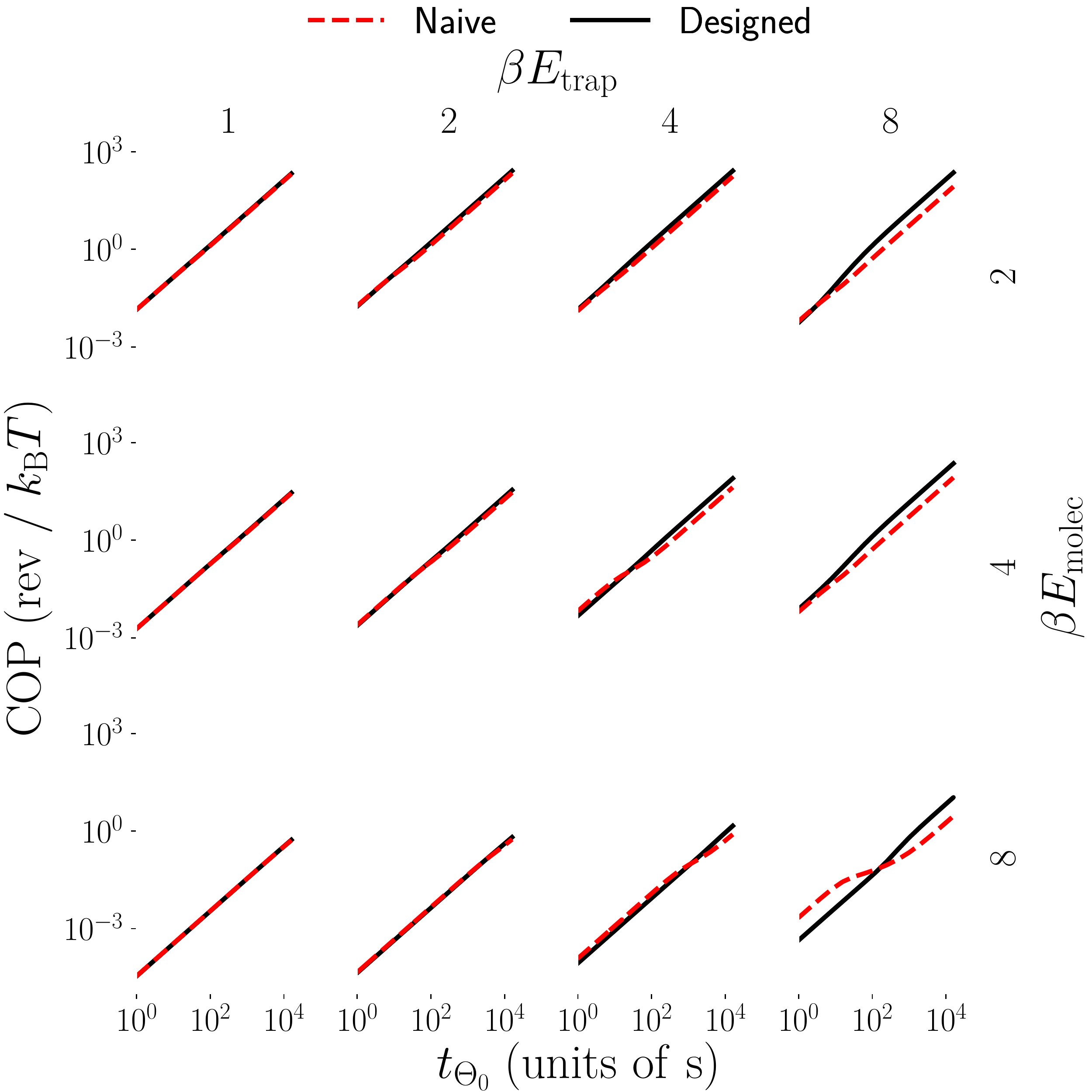} 
	}
	\caption{Coefficient of performance $\text{COP} \equiv J/W_{\rm tot}^{\rm ex}$ as a function of protocol duration $t_{\Prot}$, for naive (dashed red lines) and designed protocols (solid black lines). Same variation of $\beta E_{\rm trap}$ and $\beta E_{\rm molec}$ as in Fig.~\ref{fig:TheAutocovar}.}
	\label{fig:efficiency_plot}
\end{figure}

To investigate this tradeoff between induced flux and expended work, we quantify the coefficient of performance as $\text{COP} \equiv J / W_{\rm tot}^{\rm ex}$, where $W_{\rm tot}^{\rm ex}$ is the total excess work (the energy input), and $J$ is the flux over a full protocol and hence---given our assumption of tight coupling between mechanical motion and ATP synthesis---the machine output. Figure~\ref{fig:efficiency_plot} shows the COP for the naive and designed protocols as a function of protocol duration. 

In general, at a given mechanical barrier height and trap strength, the COP increases as a function of protocol duration: in the linear-response regime the flux $J$ doesn't vary with duration (Fig.~\ref{fig:naive_optimal_fluxes}), while the excess work $\beta W_{\rm tot}^{\rm ex}$ decreases as the inverse duration, producing an arbitrarily high COP with arbitrarily long duration. The COP on designed protocols, in particular, scales approximately linearly with protocol duration. For low trap strengths and low mechanical barrier heights (upper left subplots of Fig.~\ref{fig:efficiency_plot}), when the designed protocol closely resembles the naive, the COP for both protocols follows the approximate linear scaling with protocol duration. By contrast, in other regions of parameter space, there is a distinct crossover duration, below which the naive protocol has a higher COP than the designed protocol, and above which the opposite is true.
The crossover protocol duration decreases with trap strength and increases with barrier height.  

For the system to follow the magnetic trap in phase, it must overcome the activation energy barriers. The larger the energy barrier $\beta E_{\rm molec}$, the smaller the transition rate across the barrier, the longer the relaxation time across the barrier, and hence the larger the friction coefficient. The magnetic trap lowers the relative energy of the states and barriers near its minimum, thereby favoring system motion toward the trap minimum; hence, the stronger the magnetic trap, the greater the likelihood that the system will remain in phase with it. Therefore, in order for the system to faithfully follow the magnetic trap, the system's kinetics must be relatively fast compared to the speed of the protocol, owing to some combination of the trap being sufficiently strong to overpower the system barriers and the protocol proceeding sufficiently slowly. 

In summary, for modest-to-long protocol durations, the designed protocol remains closer to equilibrium, requiring less excess work (Fig.~\ref{fig:naive_optimal_works}), and hence achieving a greater COP.

\section{Conclusion}
Inspired by the growing interest in characterizing nonequilibrium properties of small stochastic systems, we have used a recently developed linear-response framework~\cite{SivakOptimalPath} to describe the designed driving protocols of a stochastic model system. This model system represents rotary mechanochemical molecular motors (most notably $\rm F_{1}$-ATP synthase) that couple periodic mechanical motion to progress of a chemical reaction.

We found that the driving velocity of protocols designed to minimize work can vary by orders of magnitude across a given protocol, leading to designed protocols that deviate significantly from naive protocols. In particular, a stochastic system generally dissipates less extra energy during a driven nonequilibrium process when the external driving proceeds rapidly away from mechanical energy barriers, then slows down around barriers to provide sufficient time for thermal fluctuations to kick the system over barriers with minimal energetic cost.

We tuned the operation of a rotary mechanochemical system according to a continuous and deterministic schedule that---near equilibrium---minimizes the excess work; however, unlike single-molecule experiments performed \textit{in vitro}, autonomous molecular motors---including $\Fi$-ATP synthase---are themselves driven by strongly fluctuating soft-matter objects (other proteins or protein subunits), and thus generally experience stochastic rather than deterministic driving protocols~\cite{Large2018}. Furthermore, chemical reactions occur on much faster timescales than protein conformational rearrangements, so chemical driving consists of effectively discrete control parameter jumps. Stochastic discrete protocols therefore more faithfully recreate the driving conditions of autonomous molecular machines. The intuitive implication is that to reduce energetic costs (protons run down the electrochemical potential difference across the mitochondrial membrane), $\Fo$ should position its metastable states at $\Fi$'s hopping regime, thereby utilizing thermal fluctuations during $\Fo$'s pauses to cross $\Fi$'s energetic barriers.

Unlike previous work which examined designed driving on non-periodic energetic landscapes~\cite{Sivak:2016:PhysRevE}, the periodic boundary conditions of this simple model system allow for the possibility of the protocol to finish a cycle of operation with most of the probability density remaining where it started, with the protocol effectively `lapping' the system. This 
decrease in flux also 
results in a decrease in the required excess work to perform the protocol, since most of the excess work accumulated throughout the protocol is recovered when the protocol returns to the initial state. 
This suggests that there are tradeoffs between protocol duration, energetic cost, and flux maximization that would be an interesting avenue for future research.

As energy barriers increase, force fluctuations relax more slowly, and hence standard methods for estimating the integral of the autocovariance function become increasingly computationally expensive. For future research, approximations of slowly relaxing correlation functions~\cite{Optimal5} may prove useful, and in simpler systems one can sidestep altogether the explicit calculation of the autocovariance function~\cite{zulkowskiNewMethod}.

Current experimental methods can test these general predictions: initial experiments on $\rm F_1$ would hold the magnetic trap orientation constant at different angles $\theta_0$, measure the ensuing force autocovariance function, and thereby approximate the generalized friction coefficient. Rapid rotation of the magnetic trap according to the designed schedule or alternatively with a constant angular velocity (naive) schedule could quantify the energetic savings accruing to efficient operation~\cite{TafoyaLarge2018}.

\begin{acknowledgments}
	The authors thank Steven Large, Aidan Brown, and Alexandra Kasper (SFU Physics) for insightful discussions and comments on the manuscript. This work was supported by a Natural Sciences and Engineering Research Council of Canada (NSERC) Discovery Grant (D.A.S.), a Tier-II Canada Research Chair (D.A.S.), and was enabled in part by support provided by WestGrid (\href{www.westgrid.ca}{www.westgrid.ca}) and Compute Canada Calcul Canada (\href{www.computecanada.ca}{www.computecanada.ca}).
\end{acknowledgments}

\appendix 

\section{Simulation details}
\label{sec:app_sim_details}
Calculation of the numerical solution to the master equation for a given time step is broken up into work and heat substeps. 
First, the external control parameter changes from $\theta_0(t_i)$ to $\theta_0(t_{n+1})$ according to
\begin{align}
\theta_{0}(t_{n+1}) = \theta_{0}(t_{n}) + \dfrac{\md \theta_{0}}{\md t}\Bigg|_{t = t_{n}} \Delta t\ , \label{eq:theta_update}
\end{align} 
thereby performing on the system an ensemble-averaged work, 
\begin{align}
W_{i,i+1} &= \left\langle E(E_{\rm trap},\theta_{0}(t_{n+1}))\right\rangle_{\mathbf{p}(t_{n})}\nonumber \\
&\hspace{1cm}- \left\langle E(E_{\rm trap},\theta_{0}(t_i))\right\rangle_{\mathbf{p}(t_{n})} \ ,
\end{align}
for ensemble average $\langle \cdots \rangle_{\mathbf{p}(t_{n})}$ over the instantaneous distribution 
\begin{align}
\mathbf{p}(t_{n}) \equiv \begin{bmatrix}
p(\theta_{1}|t_{n},\Prot(t_{n})) \\  
p(\theta_{2}|t_{n},\Prot(t_{n})) \\ 
p(\theta_{3}|t_{n},\Prot(t_{n})) 
\end{bmatrix}^{T} \ ,
\end{align} 
with control parameter history $\Prot(t_{n}) \equiv \{\theta_{0}(t')\}_{t'=0}^{t_{n}}$ and internal energy $E_{j}(E_{\rm trap},\theta_{0}(t_{n})) = E_{\rm trap}\cos(\theta_{j}-\theta_{0}(t_{n}))$ for state $j$ at a given control parameter $\theta_{0}(t_{n})$, all at time $t_{n}$. 

Secondly, the system relaxes at fixed control parameter value $\theta_0(t_{n+1})$. The system's dynamics evolve according to the Master Equation~\eqref{eq:master_equation},
\begin{align}
\mathbf{p}(t_{n+1}) = \mathbf{p}(t_{n}) \exp\left(\bm{\mcR}\Delta t\right) \ . 
\end{align} 
Here the time step $\Delta t \equiv t_{n+1} - t_{n}$ is chosen to be much less than the time scale over which the transition rates change significantly so that, to a good approximation, $\bm{\mcR}(t_{n+1}) = \bm{\mcR}(t_{n})=\bm{\mcR}$. 

The system probability distribution evolves over the course of a full protocol $\Prot$, described by a set of control parameter values 
\begin{align}
\Prot \equiv \Prot(t_{\Prot}) = \{\theta_{0}(t')\}_{t'=0}^{t_{\Prot}} \ .
\end{align} 
Thus control parameter $\theta_{0}$ evolves with velocity $\frac{\md \theta_{0}}{\md t}\big|_{t=t_{n}}$ at time $t_{n}$. 

We are most interested in steady-state behavior, compared to transients following some arbitrary initialization.  Thus we initialize the system by evolving from an equilibrium distribution for $\theta_{0}(0) = 0$ until it reaches a \emph{periodic steady state}, characterized by the system returning to the same distribution at identical points in its periodic cycle, as judged by total variation distance being less than some small number:
\begin{align}
\dfrac{1}{2}\sum\limits_{i=1}^{3}|p_{i}(t) - p_{i}(t + t_{\rm cycle})| < 0.00001\ ,
\end{align}
where $t_{\rm cycle}$ is one cycle duration.

We choose the integration time step size $\Delta t$ to satisfy
\begin{align}
\dfrac{\md \theta_{0}^{\rm max}}{\md t}\ \Delta t \le 10^{-2}\ .\label{eq:dt_ineq} 
\end{align} 
Since the fastest velocity we have explored in this work is $\md \theta_{0}^{\rm max}/\md t \sim 10^{3}$, we use $\Delta t = \num{e-5}$.

The total work $W$ is the sum of substep works~\cite{HeatWork}. The excess work is
\begin{align}
W_{\rm ex} \equiv W - \Delta F \ ,
\end{align} 
where
\begin{align}
\Delta F = F(\theta_{0}(t_{n+1})) - F(\theta_{0}(t_{n}))
\end{align} 
is the equilibrium free energy difference. The equilibrium free energy is  
\begin{align}
F(\theta_{0}(t_{n})) = \langle E(E_{\rm trap},\theta_{0}(t_{n})) \rangle_{\bm{\pi}(\theta_{0}(t_{n}))} - S(\theta_{0}(t_{n})) \ ,
\end{align} 
and the equilibrium entropy is
\begin{align}
S(\theta_{0}(t_{n})) = -\sum_{i}\pi(\theta_{i}|\theta_{0}(t_{n}))\ln[\pi(\theta_{i}|\theta_{0}(t_{n}))] \ .
\end{align} 
Excess work is zero for an infinitely slow protocol.

The average net probability flux for a time step $n$ is calculated from the instantaneous distribution (with evolution governed by the numerical master equation, not assuming linear-response), by summing the fluxes at each of the substeps of the calculation:
\begin{align}
3J(t_{n}) &= p(\theta_{1}|t_{n},\Prot(t_{n}))\left(\mcR_{12} - \mcR_{21}\right)\nonumber\\
&+p(\theta_{2}|t_{n},\Prot(t_{n}))\left(\mcR_{23} - \mcR_{32}\right)\nonumber\\
&+p(\theta_{3}|t_{n},\Prot(t_{n}))\left(\mcR_{31} - \mcR_{13}\right) \ .
\end{align}
The total average net probability flux sums $J(t_{n})$ over all time steps.

\section{Autocovariance calculation}
\label{sec:autocorr_calculate}

\begin{figure}[!htbp]
	\subfloat{
    	\includegraphics[clip,width=0.96\columnwidth]{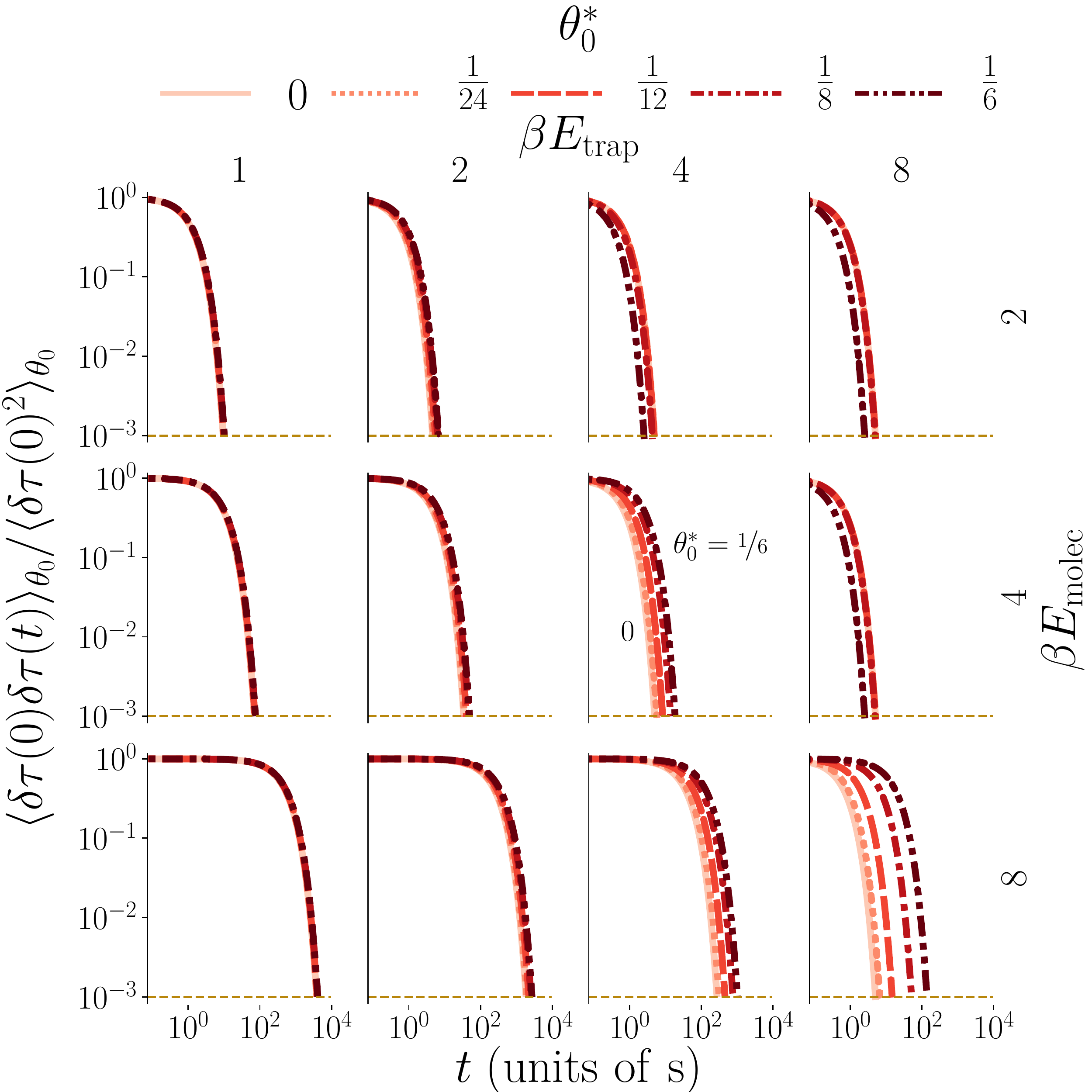}
    }
	\caption{Log-log plot of equilibrium autocorrelation $\langle\delta\tau(0)\delta\tau(t)\rangle_{\theta_{0}}$ of torque fluctuations $\delta\tau$ as a function of separation time $t$. 
	Dashed golden line denotes the calculation threshold of $\num{e-3}$. Same variation of parameters as in Fig.~\ref{fig:TheAutocovar}.
	} 
	\label{fig:correlation_plot}
\end{figure}

We calculate the autocovariance by the relation
\begin{align}
	\langle\delta\tau(0)\delta\tau(t)\rangle_{\theta_{0}} = \sum\limits_{i,j} &\pi(\theta_{i}|\theta_{0})\left[\exp(\bm{\mcR} t)\right]_{ij}\delta\tau^{i}_{\theta_{0}}\delta\tau^{j}_{\theta_{0}}\ .
\end{align}
Here $i$ and $j$ index system microstates, $\left[\exp(\bm{\mcR} t)\right]_{ij}$ give the transition probability matrix elements, $\pi(\theta_{i}|\theta_{0})$ is the equilibrium distribution, and $\delta\tau_{\theta_{0}}^{i} = \tau^{i} - \langle\tau\rangle_{\theta_{0}}$ are torque deviations from the mean, for a given control parameter $\theta_{0}$. We calculate this quantity for different times $t$ until the autocorrelation (normalized autocovariance) falls below a threshold, 
\begin{align}
	\dfrac{\langle\delta\tau(0)\delta\tau(t)\rangle_{\theta_{0}}}{\langle\delta\tau(0)^{2}\rangle_{\theta_{0}}} < \num{e-3}\ ,
\end{align}
as any further relaxation contributes negligibly to the integral in~\eqref{eq:form_friction}. Figure~\ref{fig:correlation_plot} shows that the autocorrelation function does indeed decay with time and that the relaxation timescale varies many orders of magnitude depending on $\beta E_{\rm trap}$, $\beta E_{\rm molec}$, and the magnetic field orientation $\theta_{0}^{*}$. 

\section{Setting the time scale}
\label{sec:app_time_scale}
To map simulation time to physical time, we set the physical diffusion coefficient equal to the simulation diffusion coefficient. 
The experimental rotational diffusion coefficient is approximately (according to the Stokes-Einstein relation) $D = (\beta m\gamma)^{-1}$, for inverse temperature $\beta \equiv \left(k_{\rm B}T \right)^{-1}$ and frictional drag coefficient $m\gamma_{\rm exp} = 8\pi\eta r^{3} + 6\pi\eta y^{2}r$ of a spherical bead rotationally diffusing about a point a distance $y$ from the center of the bead~\cite{Xu2008}. 
In a typical experiment, the magnetic bead is tethered to the molecular machine via a streptavidin-biotin linkage~\cite{You2012,Watanabe2013}, so the bead rotationally diffuses about its edge, giving $y = r_{\rm bead}$. 
Our calculations assume a bead with radius $r_{\rm bead} = 275\si{\nano\meter}$, within the range of typical sizes for beads of this type. 
We also use typical experimental parameters for water viscosity $\eta = 10^{-9}\;\si{\pico\newton\second\per\square\nano\meter}$ and temperature $T =300\;\si{\kelvin}$. 
We thus estimate an experimental diffusion coefficient $D_{\rm exp} \approx 4.5\ \si{\square\radian\per\second}$. 
The simulated diffusion coefficient can be calculated from the rate matrix $\bm{\mcR}$ to be $D_{\rm sim} \approx 4.4\ \rm \si{\square\radian}\,\rm (\Delta t)^{-1}$. 
Therefore, the simulation diffusion coefficient equals the estimated experimental diffusion coefficient for $\Delta t = 1\;\si{\second}$, thereby setting the simulation time scale.

\section{Comparison between designed and naive protocols}

\begin{figure}[!htbp]
	\subfloat{
    	\includegraphics[clip,width=0.96\columnwidth]{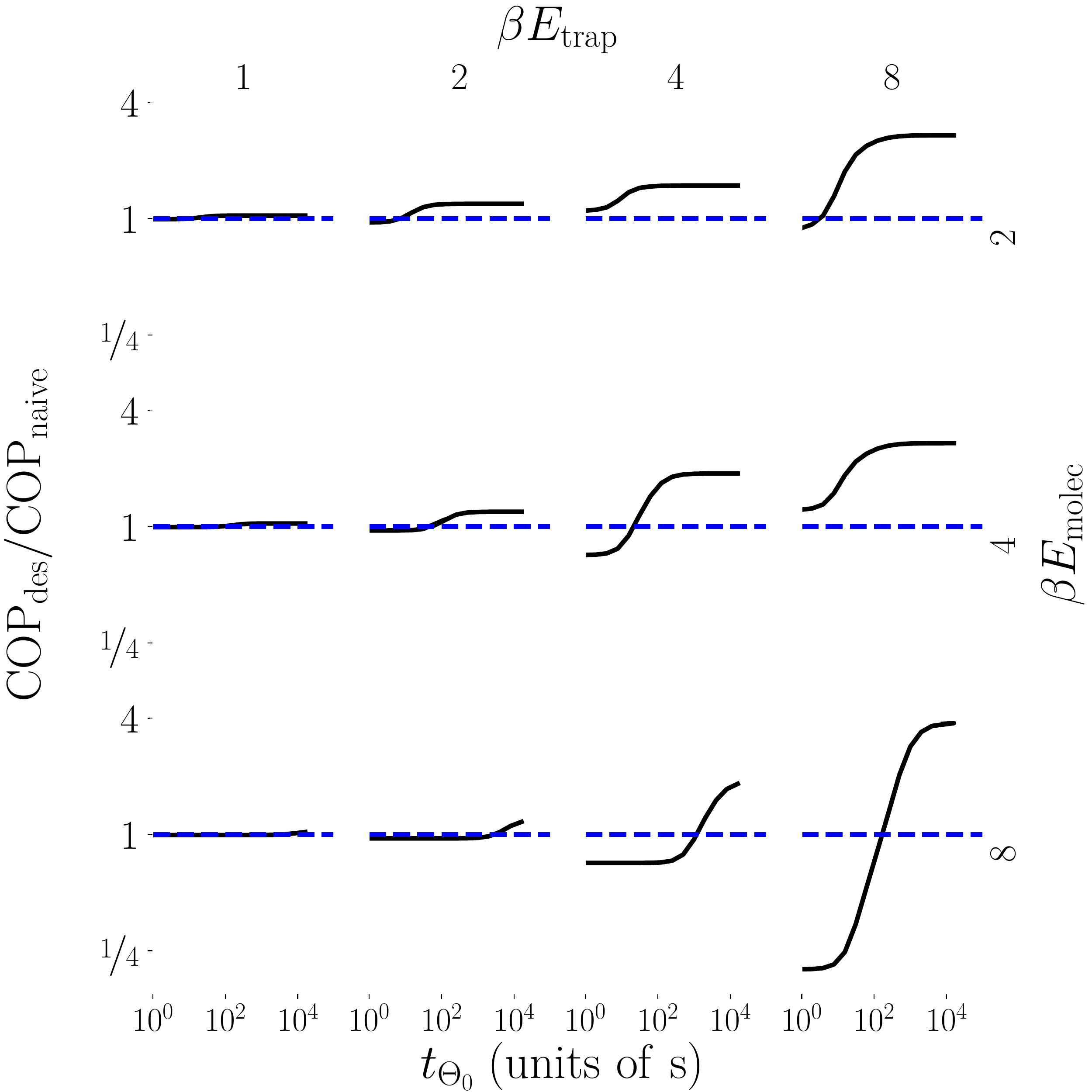}
}
    \caption{
    Ratio of designed protocol COP to naive protocol COP as a function of protocol duration $t_{\Theta_{0}}$. 
    Solid black curves show the ratio. 
    Dashed blue lines denote unity, and hence equal performance of the two protocol types. 
    Same variation of $\beta E_{\rm trap}$ and $\beta E_{\rm molec}$ as in Fig.~\ref{fig:TheAutocovar}.
    } 
    \label{fig:cop_ratio}
\end{figure}

Here we provide an alternative examination of the difference in performance of designed and naive protocols. Figure~\ref{fig:cop_ratio} shows the ratio between the designed protocol COP to the naive protocol COP. 
At long protocol duration, the designed protocol generally outperforms the naive protocol. 
Conversely, in the limit of short protocol duration, the naive protocol outperforms. 
For the most part, when the designed protocol does better, the outperformance is greater than when the naive protocol does better. 
In the limit of short protocol duration, for sufficiently weak traps relative to the barrier height, the system effectively has no time to respond to the external perturbation, and the two protocol types perform similarly. 

\section{Comparing linear-response approximation to exact results}
\label{sec:approx_to_exact}

We construct the approximations to the excess power by using the definition in~\eqref{eq:Pex}. Setting the velocity constant (naive protocol) in the approximation~\eqref{eq:Pex} gives that the structure of the excess power should match the structure of the friction coefficient. On the other hand, doing the same thing---except using the designed velocity~\eqref{eq:designed_vel} in the approximation---gives that the excess power should be constant throughout the protocol.

\begin{figure}[!htbp]
	\subfloat
	{
		\includegraphics[clip,width=0.96\columnwidth]{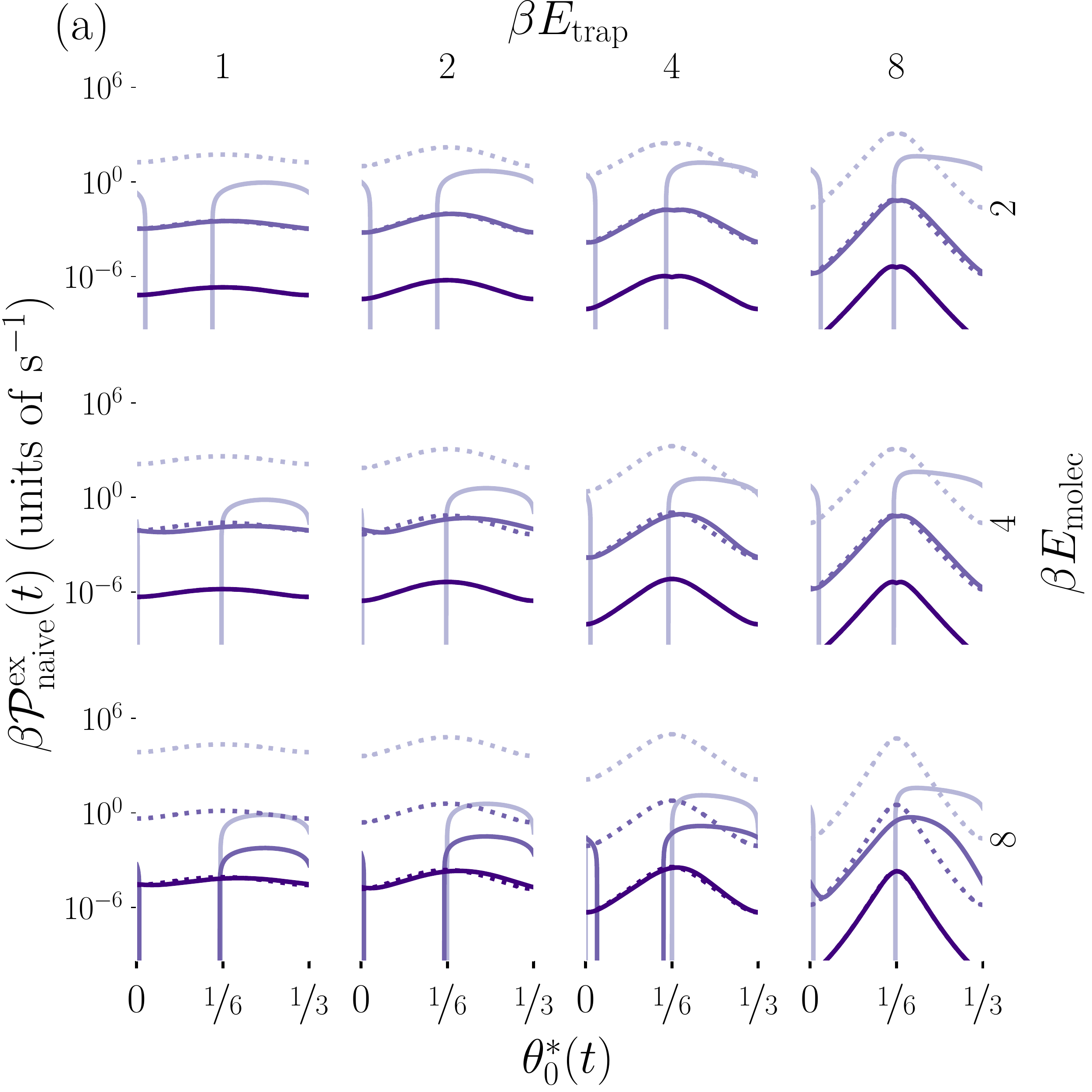} 
		\label{fig:naive_powers}
	}
	
	\subfloat
	{
		\includegraphics[clip,width=0.96\columnwidth]{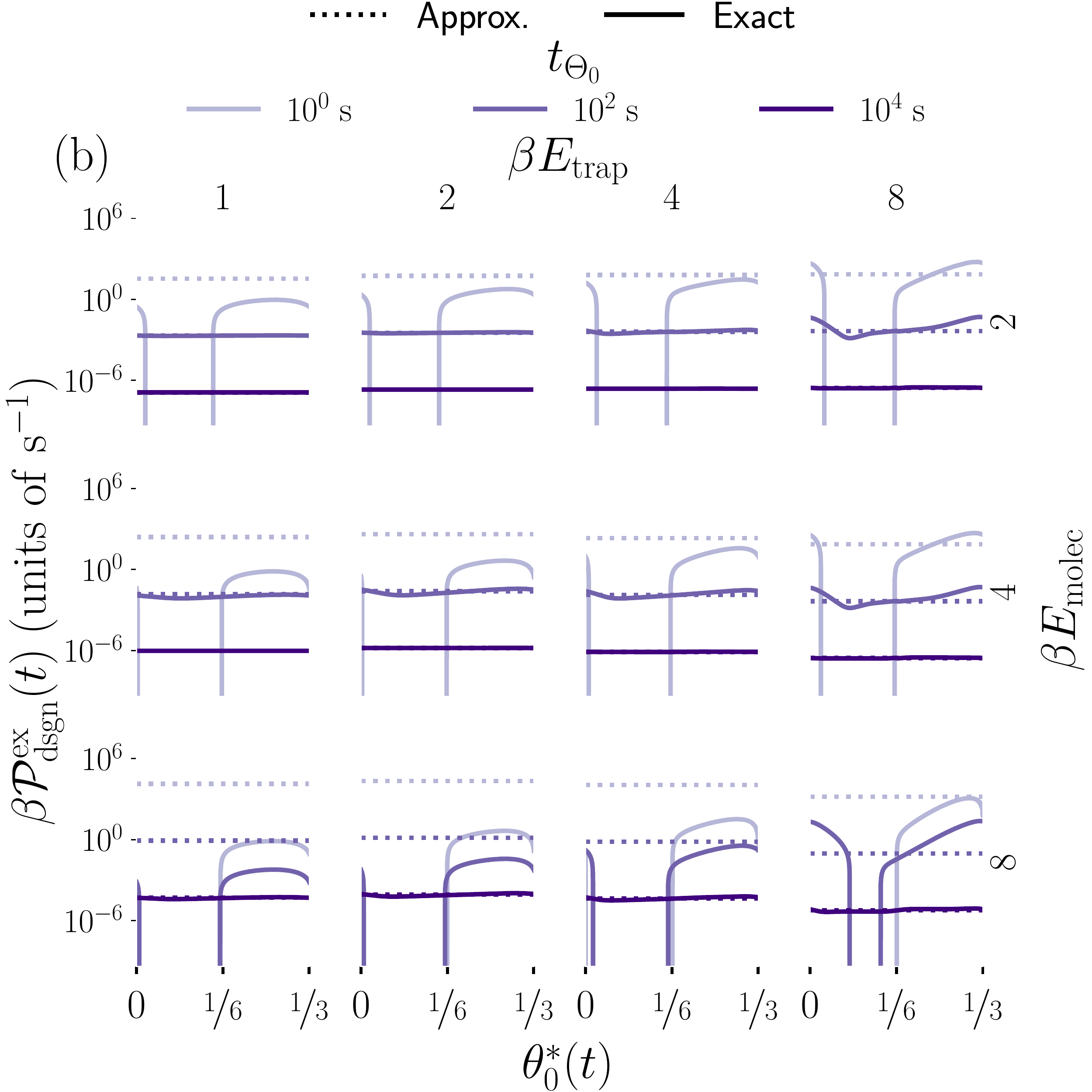} 
		\label{fig:optimal_powers}
	} 
	\caption{
	(a) Excess power $\beta \mathcal{P}^{\rm ex}_{\rm naive}$ during the naive protocol, and (b) excess power $\beta \mathcal{P}^{\rm ex}_{\rm designed}$ during the designed protocol, each as a function of instantaneous trap minimum $\theta_{0}^{*}(t)$. 
	Solid curves show excess power calculated exactly via numerical master equation solutions that do not assume linear response. 
	Dotted curves show analytical estimates using the control parameter velocity and the friction coefficient. 
	Different shades represent different protocol durations $t_{\Prot}$. 
	Same variation of $\beta E_{\rm trap}$ and $\beta E_{\rm molec}$ as in Fig.~\ref{fig:TheAutocovar}.
	}
	\label{fig:the_powers_to_approx}
\end{figure}

Figure~\ref{fig:the_powers_to_approx} shows that during both protocol types, as the protocol duration increases (driving velocity decreases), the exact system behavior converges to the approximation; however, the exact excess power disagrees with the expected result at short protocol durations. In this regime, if the protocol duration is sufficiently short and the trap strength is small relative to the height of the barrier, the exact excess power even becomes negative. 

\section{Instantaneous relative entropy}
\label{sec:inst_rel_entropy}
\begin{figure}[!ht]
	\subfloat{
		\includegraphics[clip,width=0.96\columnwidth]{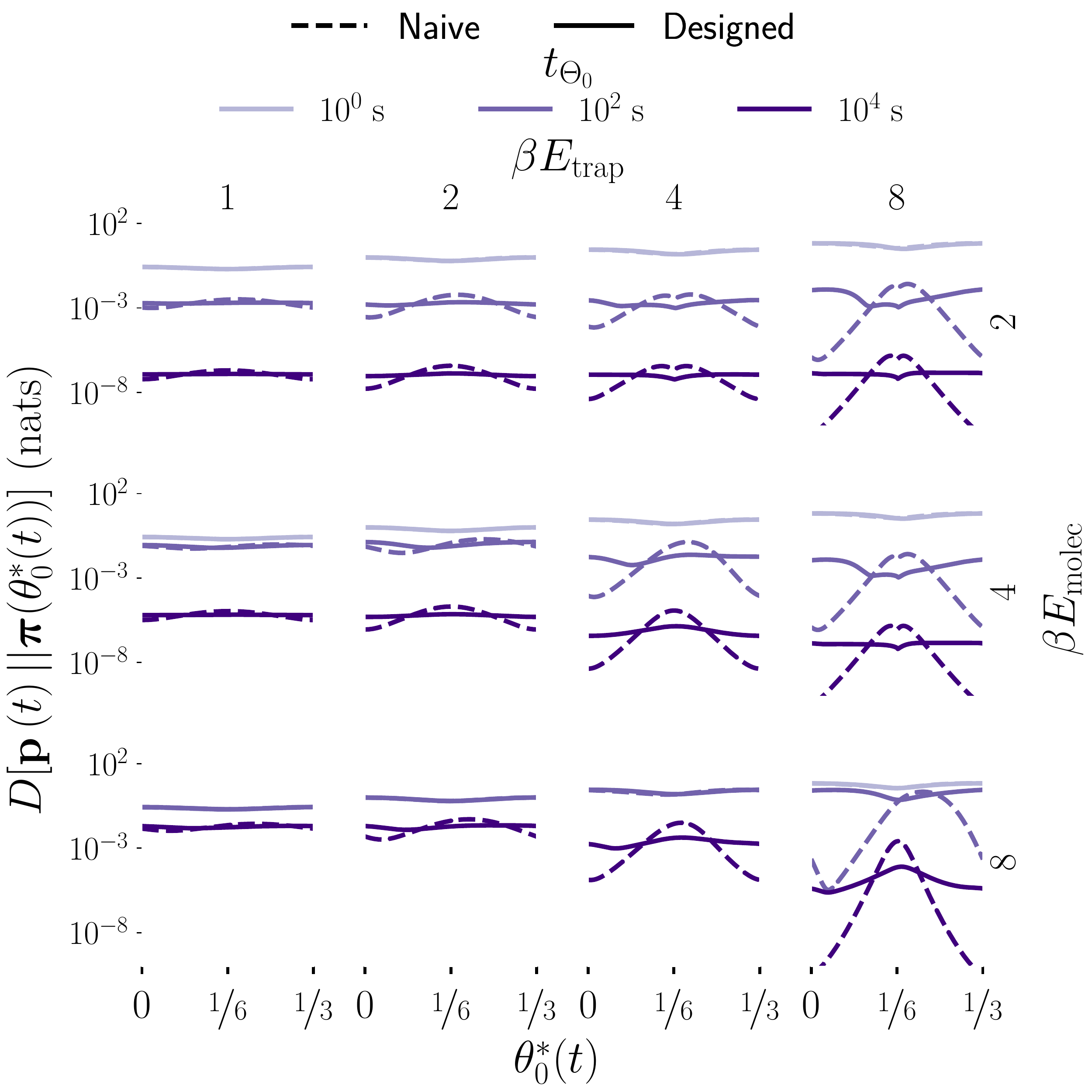} 
	}
	\caption{
	Relative entropy $D\left[\mathbf{p}(t)||\bm{\pi}(\theta_{0}^{*}(t))\right]$ between instantaneous nonequilibrium distribution $\mathbf{p}(t)$ and the equilibrium distribution $\bm{\pi}(\theta_{0}^{*}(t))$, as a function of instantaneous trap minimum $\theta_{0}^{*}(t)$, for naive (dashed lines) and designed protocols (solid lines), for various protocol durations (different shades). 
	Same variation of $\beta E_{\rm trap}$ and $\beta E_{\rm molec}$ as in Fig.~\ref{fig:TheAutocovar}.}
	\label{fig:the_distance}
\end{figure}

Figure~\ref{fig:the_distance} shows the instantaneous relative entropy as a function of instantaneous control parameter. For shorter protocols, both protocol types take the system equally far from equilibrium. With increasing protocol duration the designed protocol keeps the system a distance from equilibrium that stays relatively constant over the entire protocol. By contrast, the naive protocol produces a more variable deviation from equilibrium, with the largest deviation occurring when the trap minimum is at the barriers. 

\section{Discretization artifact in the relaxation time}
\label{sec:app_symmetry}
Our highly simplified model produces a strange phenomenon: when $E_{\rm trap}\gtrsim 2E_{\rm molec}$, the relaxation time decreases significantly at the barrier (seen in the anomalously negative slopes of some $\theta_{0}^{*} = 1/6$ autocovariance functions in Fig.~\ref{fig:autocovar}). 
When the trap minimum is at a barrier, the energies of the states flanking the barrier are equal, and the third (and only other) state, located opposite the barrier, has much higher energy and thus negligible equilibrium probability. 
For such a strong trap, the energy of the barrier equals or falls below those of the adjacent states, and thus ceases to function as a true energetic barrier. 
Since our model only resolves a small number of states and barriers, this produces an anomalously low relaxation time compared to when the trap minimum is near but not directly at the barrier. 
This, in turn, causes the friction coefficient to dip at the barrier (seen as small divots at the peak of three friction coefficient curves in Fig.~\ref{fig:friction}).
This artifact---of a decrease in relaxation time resulting in a mild decrease in friction coefficient near the barrier---vanishes for more complicated models, for example those with a continuous state space that retain barriers even when $E_{\rm trap} \gtrsim 2E_{\rm molec}$.

\FloatBarrier

\bibliographystyle{apsrev4-1}
\bibliography{Paperbib}

\begin{thebibliography}{46}%
\makeatletter
\providecommand \@ifxundefined [1]{%
 \@ifx{#1\undefined}
}%
\providecommand \@ifnum [1]{%
 \ifnum #1\expandafter \@firstoftwo
 \else \expandafter \@secondoftwo
 \fi
}%
\providecommand \@ifx [1]{%
 \ifx #1\expandafter \@firstoftwo
 \else \expandafter \@secondoftwo
 \fi
}%
\providecommand \natexlab [1]{#1}%
\providecommand \enquote  [1]{``#1''}%
\providecommand \bibnamefont  [1]{#1}%
\providecommand \bibfnamefont [1]{#1}%
\providecommand \citenamefont [1]{#1}%
\providecommand \href@noop [0]{\@secondoftwo}%
\providecommand \href [0]{\begingroup \@sanitize@url \@href}%
\providecommand \@href[1]{\@@startlink{#1}\@@href}%
\providecommand \@@href[1]{\endgroup#1\@@endlink}%
\providecommand \@sanitize@url [0]{\catcode `\\12\catcode `\$12\catcode
  `\&12\catcode `\#12\catcode `\^12\catcode `\_12\catcode `\%12\relax}%
\providecommand \@@startlink[1]{}%
\providecommand \@@endlink[0]{}%
\providecommand \url  [0]{\begingroup\@sanitize@url \@url }%
\providecommand \@url [1]{\endgroup\@href {#1}{\urlprefix }}%
\providecommand \urlprefix  [0]{URL }%
\providecommand \Eprint [0]{\href }%
\providecommand \doibase [0]{http://dx.doi.org/}%
\providecommand \selectlanguage [0]{\@gobble}%
\providecommand \bibinfo  [0]{\@secondoftwo}%
\providecommand \bibfield  [0]{\@secondoftwo}%
\providecommand \translation [1]{[#1]}%
\providecommand \BibitemOpen [0]{}%
\providecommand \bibitemStop [0]{}%
\providecommand \bibitemNoStop [0]{.\EOS\space}%
\providecommand \EOS [0]{\spacefactor3000\relax}%
\providecommand \BibitemShut  [1]{\csname bibitem#1\endcsname}%
\let\auto@bib@innerbib\@empty
\bibitem [{\citenamefont {Kolomeisky}(2013)}]{Kolomeisky2013}%
  \BibitemOpen
  \bibfield  {author} {\bibinfo {author} {\bibfnamefont {A.~B.}\ \bibnamefont
  {Kolomeisky}},\ }\href {http://stacks.iop.org/0953-8984/25/i=46/a=463101}
  {\bibfield  {journal} {\bibinfo  {journal} {J. Phys. Condens. Matter}\
  }\textbf {\bibinfo {volume} {25}},\ \bibinfo {pages} {463101} (\bibinfo
  {year} {2013})}\BibitemShut {NoStop}%
\bibitem [{\citenamefont {Schliwa}\ and\ \citenamefont
  {Woehlke}(2003)}]{Schliwa2003}%
  \BibitemOpen
  \bibfield  {author} {\bibinfo {author} {\bibfnamefont {M.}~\bibnamefont
  {Schliwa}}\ and\ \bibinfo {author} {\bibfnamefont {G.}~\bibnamefont
  {Woehlke}},\ }\href {http://dx.doi.org/10.1038/nature01601
  http://10.0.4.14/nature01601} {\bibfield  {journal} {\bibinfo  {journal}
  {Nature}\ }\textbf {\bibinfo {volume} {422}},\ \bibinfo {pages} {759}
  (\bibinfo {year} {2003})}\BibitemShut {NoStop}%
\bibitem [{\citenamefont {Yoshida}\ \emph {et~al.}(2001)\citenamefont
  {Yoshida}, \citenamefont {Muneyuki},\ and\ \citenamefont {Hisabori}}]{ATP}%
  \BibitemOpen
  \bibfield  {author} {\bibinfo {author} {\bibfnamefont {M.}~\bibnamefont
  {Yoshida}}, \bibinfo {author} {\bibfnamefont {E.}~\bibnamefont {Muneyuki}}, \
  and\ \bibinfo {author} {\bibfnamefont {T.}~\bibnamefont {Hisabori}},\ }\href
  {http://dx.doi.org/10.1038/35089509} {\bibfield  {journal} {\bibinfo
  {journal} {Nat. Rev. Mol. Cell Bio.}\ }\textbf {\bibinfo {volume} {2}},\
  \bibinfo {pages} {669} (\bibinfo {year} {2001})}\BibitemShut {NoStop}%
\bibitem [{\citenamefont {Elston}\ \emph {et~al.}(1998)\citenamefont {Elston},
  \citenamefont {Wang},\ and\ \citenamefont {Oster}}]{Coupling}%
  \BibitemOpen
  \bibfield  {author} {\bibinfo {author} {\bibfnamefont {T.}~\bibnamefont
  {Elston}}, \bibinfo {author} {\bibfnamefont {H.}~\bibnamefont {Wang}}, \ and\
  \bibinfo {author} {\bibfnamefont {G.}~\bibnamefont {Oster}},\ }\href
  {http://dx.doi.org/10.1038/35185} {\bibfield  {journal} {\bibinfo  {journal}
  {Nature}\ }\textbf {\bibinfo {volume} {391}},\ \bibinfo {pages} {510}
  (\bibinfo {year} {1998})}\BibitemShut {NoStop}%
\bibitem [{\citenamefont {Minamino}\ and\ \citenamefont
  {Imada}(2015)}]{Minamino:2015hd}%
  \BibitemOpen
  \bibfield  {author} {\bibinfo {author} {\bibfnamefont {T.}~\bibnamefont
  {Minamino}}\ and\ \bibinfo {author} {\bibfnamefont {K.}~\bibnamefont
  {Imada}},\ }\href {\doibase https://doi.org/10.1016/j.tim.2014.12.011}
  {\bibfield  {journal} {\bibinfo  {journal} {Trends Microbiol}\ }\textbf
  {\bibinfo {volume} {23}},\ \bibinfo {pages} {267 } (\bibinfo {year}
  {2015})}\BibitemShut {NoStop}%
\bibitem [{\citenamefont {Noji}\ \emph {et~al.}(1997)\citenamefont {Noji},
  \citenamefont {Yasuda}, \citenamefont {Yoshida},\ and\ \citenamefont
  {Kinosita}}]{Single-Molecule1}%
  \BibitemOpen
  \bibfield  {author} {\bibinfo {author} {\bibfnamefont {H.}~\bibnamefont
  {Noji}}, \bibinfo {author} {\bibfnamefont {R.}~\bibnamefont {Yasuda}},
  \bibinfo {author} {\bibfnamefont {M.}~\bibnamefont {Yoshida}}, \ and\
  \bibinfo {author} {\bibfnamefont {K.}~\bibnamefont {Kinosita}},\ }\href
  {http://dx.doi.org/10.1038/386299a0} {\bibfield  {journal} {\bibinfo
  {journal} {Nature}\ }\textbf {\bibinfo {volume} {386}},\ \bibinfo {pages}
  {299} (\bibinfo {year} {1997})}\BibitemShut {NoStop}%
\bibitem [{\citenamefont {Rondelez}\ \emph {et~al.}(2005)\citenamefont
  {Rondelez}, \citenamefont {Tresset}, \citenamefont {Nakashima}, \citenamefont
  {Kato-Yamada}, \citenamefont {Fujita}, \citenamefont {Takeuchi},\ and\
  \citenamefont {Noji}}]{Single-Molecule2}%
  \BibitemOpen
  \bibfield  {author} {\bibinfo {author} {\bibfnamefont {Y.}~\bibnamefont
  {Rondelez}}, \bibinfo {author} {\bibfnamefont {G.}~\bibnamefont {Tresset}},
  \bibinfo {author} {\bibfnamefont {T.}~\bibnamefont {Nakashima}}, \bibinfo
  {author} {\bibfnamefont {Y.}~\bibnamefont {Kato-Yamada}}, \bibinfo {author}
  {\bibfnamefont {H.}~\bibnamefont {Fujita}}, \bibinfo {author} {\bibfnamefont
  {S.}~\bibnamefont {Takeuchi}}, \ and\ \bibinfo {author} {\bibfnamefont
  {H.}~\bibnamefont {Noji}},\ }\href {http://dx.doi.org/10.1038/nature03277}
  {\bibfield  {journal} {\bibinfo  {journal} {Nature}\ }\textbf {\bibinfo
  {volume} {433}},\ \bibinfo {pages} {773} (\bibinfo {year}
  {2005})}\BibitemShut {NoStop}%
\bibitem [{\citenamefont {Yasuda}\ \emph {et~al.}(2001)\citenamefont {Yasuda},
  \citenamefont {Noji}, \citenamefont {Yoshida}, \citenamefont {Kinosita},\
  and\ \citenamefont {Itoh}}]{Single-Molecule3}%
  \BibitemOpen
  \bibfield  {author} {\bibinfo {author} {\bibfnamefont {R.}~\bibnamefont
  {Yasuda}}, \bibinfo {author} {\bibfnamefont {H.}~\bibnamefont {Noji}},
  \bibinfo {author} {\bibfnamefont {M.}~\bibnamefont {Yoshida}}, \bibinfo
  {author} {\bibfnamefont {K.}~\bibnamefont {Kinosita}}, \ and\ \bibinfo
  {author} {\bibfnamefont {H.}~\bibnamefont {Itoh}},\ }\href
  {http://dx.doi.org/10.1038/35073513} {\bibfield  {journal} {\bibinfo
  {journal} {Nature}\ }\textbf {\bibinfo {volume} {410}},\ \bibinfo {pages}
  {898} (\bibinfo {year} {2001})}\BibitemShut {NoStop}%
\bibitem [{\citenamefont {Ueno}\ \emph {et~al.}(2005)\citenamefont {Ueno},
  \citenamefont {Suzuki}, \citenamefont {Kinosita},\ and\ \citenamefont
  {Yoshida}}]{Ueno2005}%
  \BibitemOpen
  \bibfield  {author} {\bibinfo {author} {\bibfnamefont {H.}~\bibnamefont
  {Ueno}}, \bibinfo {author} {\bibfnamefont {T.}~\bibnamefont {Suzuki}},
  \bibinfo {author} {\bibfnamefont {K.}~\bibnamefont {Kinosita}}, \ and\
  \bibinfo {author} {\bibfnamefont {M.}~\bibnamefont {Yoshida}},\ }\href
  {\doibase 10.1073/pnas.0407857102} {\bibfield  {journal} {\bibinfo  {journal}
  {PNAS}\ }\textbf {\bibinfo {volume} {102}},\ \bibinfo {pages} {1333}
  (\bibinfo {year} {2005})}\BibitemShut {NoStop}%
\bibitem [{\citenamefont {Watanabe}\ \emph {et~al.}(2011)\citenamefont
  {Watanabe}, \citenamefont {Okuno}, \citenamefont {Sakakihara}, \citenamefont
  {Shimabukuro}, \citenamefont {Iino}, \citenamefont {Yoshida},\ and\
  \citenamefont {Noji}}]{Watanabe2011}%
  \BibitemOpen
  \bibfield  {author} {\bibinfo {author} {\bibfnamefont {R.}~\bibnamefont
  {Watanabe}}, \bibinfo {author} {\bibfnamefont {D.}~\bibnamefont {Okuno}},
  \bibinfo {author} {\bibfnamefont {S.}~\bibnamefont {Sakakihara}}, \bibinfo
  {author} {\bibfnamefont {K.}~\bibnamefont {Shimabukuro}}, \bibinfo {author}
  {\bibfnamefont {R.}~\bibnamefont {Iino}}, \bibinfo {author} {\bibfnamefont
  {M.}~\bibnamefont {Yoshida}}, \ and\ \bibinfo {author} {\bibfnamefont
  {H.}~\bibnamefont {Noji}},\ }\href {http://dx.doi.org/10.1038/nchembio.715}
  {\bibfield  {journal} {\bibinfo  {journal} {Nat. Chem. Biol.}\ }\textbf
  {\bibinfo {volume} {8}},\ \bibinfo {pages} {86 EP } (\bibinfo {year}
  {2011})}\BibitemShut {NoStop}%
\bibitem [{\citenamefont {Saita}\ \emph {et~al.}(2015)\citenamefont {Saita},
  \citenamefont {Suzuki}, \citenamefont {Kinosita},\ and\ \citenamefont
  {Yoshida}}]{Saita2015}%
  \BibitemOpen
  \bibfield  {author} {\bibinfo {author} {\bibfnamefont {E.-i.}\ \bibnamefont
  {Saita}}, \bibinfo {author} {\bibfnamefont {T.}~\bibnamefont {Suzuki}},
  \bibinfo {author} {\bibfnamefont {K.}~\bibnamefont {Kinosita}}, \ and\
  \bibinfo {author} {\bibfnamefont {M.}~\bibnamefont {Yoshida}},\ }\href
  {\doibase 10.1073/pnas.1422885112} {\bibfield  {journal} {\bibinfo  {journal}
  {PNAS}\ }\textbf {\bibinfo {volume} {112}},\ \bibinfo {pages} {9626}
  (\bibinfo {year} {2015})}\BibitemShut {NoStop}%
\bibitem [{\citenamefont {Rolfe}\ and\ \citenamefont
  {Brown}(1997)}]{Rolfe:1997vq}%
  \BibitemOpen
  \bibfield  {author} {\bibinfo {author} {\bibfnamefont {D.~F.}\ \bibnamefont
  {Rolfe}}\ and\ \bibinfo {author} {\bibfnamefont {G.~C.}\ \bibnamefont
  {Brown}},\ }\href {\doibase 10.1152/physrev.1997.77.3.731} {\bibfield
  {journal} {\bibinfo  {journal} {Physiol. Rev.}\ }\textbf {\bibinfo {volume}
  {77}},\ \bibinfo {pages} {731} (\bibinfo {year} {1997})}\BibitemShut
  {NoStop}%
\bibitem [{\citenamefont {Yasuda}\ \emph {et~al.}(1998)\citenamefont {Yasuda},
  \citenamefont {Noji}, \citenamefont {Kinosita},\ and\ \citenamefont
  {Yoshida}}]{ATP-Efficiency}%
  \BibitemOpen
  \bibfield  {author} {\bibinfo {author} {\bibfnamefont {R.}~\bibnamefont
  {Yasuda}}, \bibinfo {author} {\bibfnamefont {H.}~\bibnamefont {Noji}},
  \bibinfo {author} {\bibfnamefont {K.}~\bibnamefont {Kinosita}}, \ and\
  \bibinfo {author} {\bibfnamefont {M.}~\bibnamefont {Yoshida}},\ }\href
  {http://dx.doi.org/10.1016/S0092-8674(00)81456-7} {\bibfield  {journal}
  {\bibinfo  {journal} {Cell}\ }\textbf {\bibinfo {volume} {93}},\ \bibinfo
  {pages} {1117} (\bibinfo {year} {1998})}\BibitemShut {NoStop}%
\bibitem [{\citenamefont {Feynman}\ \emph {et~al.}(1966)\citenamefont
  {Feynman}, \citenamefont {Leighton},\ and\ \citenamefont
  {Sands}}]{feynman66}%
  \BibitemOpen
  \bibfield  {author} {\bibinfo {author} {\bibfnamefont {R.~P.}\ \bibnamefont
  {Feynman}}, \bibinfo {author} {\bibfnamefont {R.~B.}\ \bibnamefont
  {Leighton}}, \ and\ \bibinfo {author} {\bibfnamefont {M.}~\bibnamefont
  {Sands}},\ }\href@noop {} {\emph {\bibinfo {title} {{The Feynman Lectures on
  Physics}}}}\ (\bibinfo  {publisher} {Addison-Wesley},\ \bibinfo {year}
  {1966})\BibitemShut {NoStop}%
\bibitem [{\citenamefont {Jarzynski}\ and\ \citenamefont
  {Mazonka}(1999)}]{jarzynski99}%
  \BibitemOpen
  \bibfield  {author} {\bibinfo {author} {\bibfnamefont {C.}~\bibnamefont
  {Jarzynski}}\ and\ \bibinfo {author} {\bibfnamefont {O.}~\bibnamefont
  {Mazonka}},\ }\href {\doibase 10.1103/PhysRevE.59.6448} {\bibfield  {journal}
  {\bibinfo  {journal} {Phys. Rev. E}\ }\textbf {\bibinfo {volume} {59}},\
  \bibinfo {pages} {6448} (\bibinfo {year} {1999})}\BibitemShut {NoStop}%
\bibitem [{\citenamefont {Isojima}\ \emph {et~al.}(2016)\citenamefont
  {Isojima}, \citenamefont {Iino}, \citenamefont {Niitani}, \citenamefont
  {Noji},\ and\ \citenamefont {Tomishige}}]{Backward}%
  \BibitemOpen
  \bibfield  {author} {\bibinfo {author} {\bibfnamefont {H.}~\bibnamefont
  {Isojima}}, \bibinfo {author} {\bibfnamefont {R.}~\bibnamefont {Iino}},
  \bibinfo {author} {\bibfnamefont {Y.}~\bibnamefont {Niitani}}, \bibinfo
  {author} {\bibfnamefont {H.}~\bibnamefont {Noji}}, \ and\ \bibinfo {author}
  {\bibfnamefont {M.}~\bibnamefont {Tomishige}},\ }\href
  {http://www.nature.com/nchembio/journal/v12/n4/abs/nchembio.2028.html}
  {\bibfield  {journal} {\bibinfo  {journal} {Nat. Chem. Biol.}\ }\textbf
  {\bibinfo {volume} {12}},\ \bibinfo {pages} {290} (\bibinfo {year}
  {2016})}\BibitemShut {NoStop}%
\bibitem [{\citenamefont {Van~Delden}\ \emph {et~al.}(2005)\citenamefont
  {Van~Delden}, \citenamefont {Ter~Wiel}, \citenamefont {Pollard},
  \citenamefont {Vicario}, \citenamefont {Koumura},\ and\ \citenamefont
  {Feringa}}]{Harness}%
  \BibitemOpen
  \bibfield  {author} {\bibinfo {author} {\bibfnamefont {R.~A.}\ \bibnamefont
  {Van~Delden}}, \bibinfo {author} {\bibfnamefont {M.~K.}\ \bibnamefont
  {Ter~Wiel}}, \bibinfo {author} {\bibfnamefont {M.~M.}\ \bibnamefont
  {Pollard}}, \bibinfo {author} {\bibfnamefont {J.}~\bibnamefont {Vicario}},
  \bibinfo {author} {\bibfnamefont {N.}~\bibnamefont {Koumura}}, \ and\
  \bibinfo {author} {\bibfnamefont {B.~L.}\ \bibnamefont {Feringa}},\ }\href
  {http://dx.doi.org/10.1038/nature04127} {\bibfield  {journal} {\bibinfo
  {journal} {Nature}\ }\textbf {\bibinfo {volume} {437}},\ \bibinfo {pages}
  {1337} (\bibinfo {year} {2005})}\BibitemShut {NoStop}%
\bibitem [{\citenamefont {Sivak}\ and\ \citenamefont
  {Crooks}(2012{\natexlab{a}})}]{SivakOptimalPath}%
  \BibitemOpen
  \bibfield  {author} {\bibinfo {author} {\bibfnamefont {D.~A.}\ \bibnamefont
  {Sivak}}\ and\ \bibinfo {author} {\bibfnamefont {G.~E.}\ \bibnamefont
  {Crooks}},\ }\href {https://doi.org/10.1103/PhysRevLett.108.190602}
  {\bibfield  {journal} {\bibinfo  {journal} {Phys. Rev. Lett.}\ }\textbf
  {\bibinfo {volume} {108}},\ \bibinfo {pages} {190602} (\bibinfo {year}
  {2012}{\natexlab{a}})}\BibitemShut {NoStop}%
\bibitem [{\citenamefont {Mandal}\ and\ \citenamefont
  {Jarzynski}(2016)}]{MandalOptimal}%
  \BibitemOpen
  \bibfield  {author} {\bibinfo {author} {\bibfnamefont {D.}~\bibnamefont
  {Mandal}}\ and\ \bibinfo {author} {\bibfnamefont {C.}~\bibnamefont
  {Jarzynski}},\ }\href {http://stacks.iop.org/1742-5468/2016/i=6/a=063204}
  {\bibfield  {journal} {\bibinfo  {journal} {J. Stat. Mech.}\ }\textbf
  {\bibinfo {volume} {2016}},\ \bibinfo {pages} {063204} (\bibinfo {year}
  {2016})}\BibitemShut {NoStop}%
\bibitem [{\citenamefont {Zulkowski}\ \emph {et~al.}(2012)\citenamefont
  {Zulkowski}, \citenamefont {Sivak}, \citenamefont {Crooks},\ and\
  \citenamefont {DeWeese}}]{Optimal1}%
  \BibitemOpen
  \bibfield  {author} {\bibinfo {author} {\bibfnamefont {P.~R.}\ \bibnamefont
  {Zulkowski}}, \bibinfo {author} {\bibfnamefont {D.~A.}\ \bibnamefont
  {Sivak}}, \bibinfo {author} {\bibfnamefont {G.~E.}\ \bibnamefont {Crooks}}, \
  and\ \bibinfo {author} {\bibfnamefont {M.~R.}\ \bibnamefont {DeWeese}},\
  }\href {https://doi.org/10.1103/PhysRevE.86.041148} {\bibfield  {journal}
  {\bibinfo  {journal} {Phys. Rev. E.}\ }\textbf {\bibinfo {volume} {86}},\
  \bibinfo {pages} {041148} (\bibinfo {year} {2012})}\BibitemShut {NoStop}%
\bibitem [{\citenamefont {Zulkowski}\ \emph {et~al.}(2013)\citenamefont
  {Zulkowski}, \citenamefont {Sivak},\ and\ \citenamefont
  {DeWeese}}]{Optimal2}%
  \BibitemOpen
  \bibfield  {author} {\bibinfo {author} {\bibfnamefont {P.~R.}\ \bibnamefont
  {Zulkowski}}, \bibinfo {author} {\bibfnamefont {D.~A.}\ \bibnamefont
  {Sivak}}, \ and\ \bibinfo {author} {\bibfnamefont {M.~R.}\ \bibnamefont
  {DeWeese}},\ }\href {http://dx.doi.org/10.1371/journal.pone.0082754}
  {\bibfield  {journal} {\bibinfo  {journal} {PLoS One}\ }\textbf {\bibinfo
  {volume} {8}},\ \bibinfo {pages} {e82754} (\bibinfo {year}
  {2013})}\BibitemShut {NoStop}%
\bibitem [{\citenamefont {Gomez-Marin}\ \emph {et~al.}(2008)\citenamefont
  {Gomez-Marin}, \citenamefont {Schmiedl},\ and\ \citenamefont
  {Seifert}}]{Optimal3}%
  \BibitemOpen
  \bibfield  {author} {\bibinfo {author} {\bibfnamefont {A.}~\bibnamefont
  {Gomez-Marin}}, \bibinfo {author} {\bibfnamefont {T.}~\bibnamefont
  {Schmiedl}}, \ and\ \bibinfo {author} {\bibfnamefont {U.}~\bibnamefont
  {Seifert}},\ }\href {http://dx.doi.org/10.1063/1.2948948} {\bibfield
  {journal} {\bibinfo  {journal} {J. Chem. Phys.}\ }\textbf {\bibinfo {volume}
  {129}},\ \bibinfo {pages} {024114} (\bibinfo {year} {2008})}\BibitemShut
  {NoStop}%
\bibitem [{\citenamefont {Schmiedl}\ and\ \citenamefont
  {Seifert}(2007)}]{Optimal4}%
  \BibitemOpen
  \bibfield  {author} {\bibinfo {author} {\bibfnamefont {T.}~\bibnamefont
  {Schmiedl}}\ and\ \bibinfo {author} {\bibfnamefont {U.}~\bibnamefont
  {Seifert}},\ }\href {https://doi.org/10.1103/PhysRevLett.98.108301}
  {\bibfield  {journal} {\bibinfo  {journal} {Phys. Rev. Lett.}\ }\textbf
  {\bibinfo {volume} {98}},\ \bibinfo {pages} {108301} (\bibinfo {year}
  {2007})}\BibitemShut {NoStop}%
\bibitem [{\citenamefont {Bonan{\c{c}}a}\ and\ \citenamefont
  {Deffner}(2014)}]{Optimal5}%
  \BibitemOpen
  \bibfield  {author} {\bibinfo {author} {\bibfnamefont {M.~V.}\ \bibnamefont
  {Bonan{\c{c}}a}}\ and\ \bibinfo {author} {\bibfnamefont {S.}~\bibnamefont
  {Deffner}},\ }\href {http://dx.doi.org/10.1063/1.4885277} {\bibfield
  {journal} {\bibinfo  {journal} {J. Chem. Phys.}\ }\textbf {\bibinfo {volume}
  {140}},\ \bibinfo {pages} {244119} (\bibinfo {year} {2014})}\BibitemShut
  {NoStop}%
\bibitem [{\citenamefont {Sivak}\ and\ \citenamefont
  {Crooks}(2016)}]{Sivak:2016:PhysRevE}%
  \BibitemOpen
  \bibfield  {author} {\bibinfo {author} {\bibfnamefont {D.~A.}\ \bibnamefont
  {Sivak}}\ and\ \bibinfo {author} {\bibfnamefont {G.~E.}\ \bibnamefont
  {Crooks}},\ }\href {\doibase 10.1103/PhysRevE.94.052106} {\bibfield
  {journal} {\bibinfo  {journal} {Phys. Rev. E}\ }\textbf {\bibinfo {volume}
  {94}},\ \bibinfo {pages} {052106} (\bibinfo {year} {2016})}\BibitemShut
  {NoStop}%
\bibitem [{\citenamefont {Itoh}\ \emph {et~al.}(2004)\citenamefont {Itoh},
  \citenamefont {Takahashi}, \citenamefont {Adachi}, \citenamefont {Noji},
  \citenamefont {Yasuda}, \citenamefont {Yoshida},\ and\ \citenamefont
  {Kinosita}}]{Mechanically_Driven}%
  \BibitemOpen
  \bibfield  {author} {\bibinfo {author} {\bibfnamefont {H.}~\bibnamefont
  {Itoh}}, \bibinfo {author} {\bibfnamefont {A.}~\bibnamefont {Takahashi}},
  \bibinfo {author} {\bibfnamefont {K.}~\bibnamefont {Adachi}}, \bibinfo
  {author} {\bibfnamefont {H.}~\bibnamefont {Noji}}, \bibinfo {author}
  {\bibfnamefont {R.}~\bibnamefont {Yasuda}}, \bibinfo {author} {\bibfnamefont
  {M.}~\bibnamefont {Yoshida}}, \ and\ \bibinfo {author} {\bibfnamefont
  {K.}~\bibnamefont {Kinosita}},\ }\href
  {http://dx.doi.org/10.1038/nature02212} {\bibfield  {journal} {\bibinfo
  {journal} {Nature}\ }\textbf {\bibinfo {volume} {427}},\ \bibinfo {pages}
  {465} (\bibinfo {year} {2004})}\BibitemShut {NoStop}%
\bibitem [{\citenamefont {Valentine}\ \emph {et~al.}(2006)\citenamefont
  {Valentine}, \citenamefont {Fordyce}, \citenamefont {Krzysiak}, \citenamefont
  {Gilbert},\ and\ \citenamefont {Block}}]{Valentine2006}%
  \BibitemOpen
  \bibfield  {author} {\bibinfo {author} {\bibfnamefont {M.~T.}\ \bibnamefont
  {Valentine}}, \bibinfo {author} {\bibfnamefont {P.~M.}\ \bibnamefont
  {Fordyce}}, \bibinfo {author} {\bibfnamefont {T.~C.}\ \bibnamefont
  {Krzysiak}}, \bibinfo {author} {\bibfnamefont {S.~P.}\ \bibnamefont
  {Gilbert}}, \ and\ \bibinfo {author} {\bibfnamefont {S.~M.}\ \bibnamefont
  {Block}},\ }\href {http://dx.doi.org/10.1038/ncb1394 http://10.0.4.14/ncb1394
  https://www.nature.com/articles/ncb1394{\#}supplementary-information}
  {\bibfield  {journal} {\bibinfo  {journal} {Nat. Cell Biol.}\ }\textbf
  {\bibinfo {volume} {8}},\ \bibinfo {pages} {470} (\bibinfo {year}
  {2006})}\BibitemShut {NoStop}%
\bibitem [{\citenamefont {Rotskoff}\ and\ \citenamefont
  {Crooks}(2015)}]{Ising}%
  \BibitemOpen
  \bibfield  {author} {\bibinfo {author} {\bibfnamefont {G.~M.}\ \bibnamefont
  {Rotskoff}}\ and\ \bibinfo {author} {\bibfnamefont {G.~E.}\ \bibnamefont
  {Crooks}},\ }\href {https://doi.org/10.1103/PhysRevE.92.060102} {\bibfield
  {journal} {\bibinfo  {journal} {Phys. Rev. E.}\ }\textbf {\bibinfo {volume}
  {92}},\ \bibinfo {pages} {060102} (\bibinfo {year} {2015})}\BibitemShut
  {NoStop}%
\bibitem [{\citenamefont {Aldous}\ and\ \citenamefont
  {Fill}(2002)}]{Aldous2002}%
  \BibitemOpen
  \bibfield  {author} {\bibinfo {author} {\bibfnamefont {D.}~\bibnamefont
  {Aldous}}\ and\ \bibinfo {author} {\bibfnamefont {J.~A.}\ \bibnamefont
  {Fill}},\ }\href@noop {} {\enquote {\bibinfo {title} {Reversible markov
  chains and random walks on graphs},}\ } (\bibinfo {year} {2002}),\ \bibinfo
  {note} {unfinished monograph, recompiled 2014, available at
  \url{http://www.stat.berkeley.edu/$\sim$aldous/RWG/book.html}}\BibitemShut
  {NoStop}%
\bibitem [{\citenamefont {van Kampen}(2007)}]{vanKampen}%
  \BibitemOpen
  \bibfield  {author} {\bibinfo {author} {\bibfnamefont {N.}~\bibnamefont {van
  Kampen}},\ }\href@noop {} {\emph {\bibinfo {title} {Stochastic Processes in
  Physics and Chemistry}}},\ \bibinfo {edition} {3rd}\ ed.\ (\bibinfo
  {publisher} {North Holland},\ \bibinfo {year} {2007})\BibitemShut {NoStop}%
\bibitem [{\citenamefont {Jacobs}(2010)}]{Jacobs2010}%
  \BibitemOpen
  \bibfield  {author} {\bibinfo {author} {\bibfnamefont {K.}~\bibnamefont
  {Jacobs}},\ }\href {\doibase 10.1017/CBO9780511815980} {\emph {\bibinfo
  {title} {Stochastic Processes for Physicists: Understanding Noisy Systems}}}\
  (\bibinfo  {publisher} {Cambridge University Press},\ \bibinfo {year}
  {2010})\BibitemShut {NoStop}%
\bibitem [{\citenamefont {You}\ \emph {et~al.}(2012)\citenamefont {You},
  \citenamefont {Iino}, \citenamefont {Watanabe},\ and\ \citenamefont
  {Noji}}]{You2012}%
  \BibitemOpen
  \bibfield  {author} {\bibinfo {author} {\bibfnamefont {H.}~\bibnamefont
  {You}}, \bibinfo {author} {\bibfnamefont {R.}~\bibnamefont {Iino}}, \bibinfo
  {author} {\bibfnamefont {R.}~\bibnamefont {Watanabe}}, \ and\ \bibinfo
  {author} {\bibfnamefont {H.}~\bibnamefont {Noji}},\ }\href {\doibase
  10.1093/nar/gks651} {\bibfield  {journal} {\bibinfo  {journal} {Nucleic Acids
  Res.}\ }\textbf {\bibinfo {volume} {40}},\ \bibinfo {pages} {e151} (\bibinfo
  {year} {2012})}\BibitemShut {NoStop}%
\bibitem [{\citenamefont {Xu}(2008)}]{Xu2008}%
  \BibitemOpen
  \bibfield  {author} {\bibinfo {author} {\bibfnamefont {L.}~\bibnamefont
  {Xu}},\ }\href {\doibase https://doi.org/10.1016/j.bbabio.2008.08.010}
  {\bibfield  {journal} {\bibinfo  {journal} {Bio. et Biophys. Acta (BBA) -
  Bioenerg.}\ }\textbf {\bibinfo {volume} {1777}},\ \bibinfo {pages} {1422 }
  (\bibinfo {year} {2008})}\BibitemShut {NoStop}%
\bibitem [{\citenamefont {Okazaki}\ and\ \citenamefont
  {Hummer}(2015)}]{DiffusionEst}%
  \BibitemOpen
  \bibfield  {author} {\bibinfo {author} {\bibfnamefont {K.}~\bibnamefont
  {Okazaki}}\ and\ \bibinfo {author} {\bibfnamefont {G.}~\bibnamefont
  {Hummer}},\ }\href {http://www.pnas.org/content/112/34/10720.short}
  {\bibfield  {journal} {\bibinfo  {journal} {Proc. Natl. Acad. Sci. USA.}\
  }\textbf {\bibinfo {volume} {112}},\ \bibinfo {pages} {10720} (\bibinfo
  {year} {2015})}\BibitemShut {NoStop}%
\bibitem [{\citenamefont {Sekimoto}\ and\ \citenamefont
  {Sasa}(1997)}]{SekimotoSasa1997}%
  \BibitemOpen
  \bibfield  {author} {\bibinfo {author} {\bibfnamefont {K.}~\bibnamefont
  {Sekimoto}}\ and\ \bibinfo {author} {\bibfnamefont {S.}~\bibnamefont
  {Sasa}},\ }\href {\doibase 10.1143/JPSJ.66.3326} {\bibfield  {journal}
  {\bibinfo  {journal} {J. Phys. Soc. Jpn}\ }\textbf {\bibinfo {volume} {66}},\
  \bibinfo {pages} {3326} (\bibinfo {year} {1997})}\BibitemShut {NoStop}%
\bibitem [{\citenamefont {Crooks}(2007)}]{Crooks2007}%
  \BibitemOpen
  \bibfield  {author} {\bibinfo {author} {\bibfnamefont {G.~E.}\ \bibnamefont
  {Crooks}},\ }\href {\doibase 10.1103/PhysRevLett.99.100602} {\bibfield
  {journal} {\bibinfo  {journal} {Phys. Rev. Lett.}\ }\textbf {\bibinfo
  {volume} {99}},\ \bibinfo {pages} {100602} (\bibinfo {year}
  {2007})}\BibitemShut {NoStop}%
\bibitem [{\citenamefont {Brown}\ and\ \citenamefont
  {Sivak}(2017)}]{Brown2017}%
  \BibitemOpen
  \bibfield  {author} {\bibinfo {author} {\bibfnamefont {A.~I.}\ \bibnamefont
  {Brown}}\ and\ \bibinfo {author} {\bibfnamefont {D.~A.}\ \bibnamefont
  {Sivak}},\ }\href {\doibase 10.1073/pnas.1707534114} {\bibfield  {journal}
  {\bibinfo  {journal} {Proc. Natl. Acad. Sci. USA.}\ }\textbf {\bibinfo
  {volume} {114}},\ \bibinfo {pages} {11057} (\bibinfo {year}
  {2017})}\BibitemShut {NoStop}%
\bibitem [{\citenamefont {Anandakrishnan}\ \emph {et~al.}(2016)\citenamefont
  {Anandakrishnan}, \citenamefont {Zhang}, \citenamefont {Donovan-Maiye},\ and\
  \citenamefont {Zuckerman}}]{Anandakrishnan2016}%
  \BibitemOpen
  \bibfield  {author} {\bibinfo {author} {\bibfnamefont {R.}~\bibnamefont
  {Anandakrishnan}}, \bibinfo {author} {\bibfnamefont {Z.}~\bibnamefont
  {Zhang}}, \bibinfo {author} {\bibfnamefont {R.}~\bibnamefont
  {Donovan-Maiye}}, \ and\ \bibinfo {author} {\bibfnamefont {D.~M.}\
  \bibnamefont {Zuckerman}},\ }\href {\doibase 10.1073/pnas.1608533113}
  {\bibfield  {journal} {\bibinfo  {journal} {PNAS}\ }\textbf {\bibinfo
  {volume} {113}},\ \bibinfo {pages} {11220} (\bibinfo {year}
  {2016})}\BibitemShut {NoStop}%
\bibitem [{\citenamefont {Cover}\ and\ \citenamefont
  {Thomas}(2006)}]{Cover:2006:Book}%
  \BibitemOpen
  \bibfield  {author} {\bibinfo {author} {\bibfnamefont {T.~M.}\ \bibnamefont
  {Cover}}\ and\ \bibinfo {author} {\bibfnamefont {J.~A.}\ \bibnamefont
  {Thomas}},\ }\href@noop {} {\emph {\bibinfo {title} {Elements of Information
  Theory}}},\ \bibinfo {edition} {2nd}\ ed.\ (\bibinfo  {publisher}
  {Wiley-Interscience},\ \bibinfo {address} {Hoboken},\ \bibinfo {year}
  {2006})\BibitemShut {NoStop}%
\bibitem [{\citenamefont {Sivak}\ and\ \citenamefont
  {Crooks}(2012{\natexlab{b}})}]{SivakCrooks2012}%
  \BibitemOpen
  \bibfield  {author} {\bibinfo {author} {\bibfnamefont {D.~A.}\ \bibnamefont
  {Sivak}}\ and\ \bibinfo {author} {\bibfnamefont {G.~E.}\ \bibnamefont
  {Crooks}},\ }\href {\doibase 10.1103/PhysRevLett.108.150601} {\bibfield
  {journal} {\bibinfo  {journal} {Phys. Rev. Lett.}\ }\textbf {\bibinfo
  {volume} {108}},\ \bibinfo {pages} {150601} (\bibinfo {year}
  {2012}{\natexlab{b}})}\BibitemShut {NoStop}%
\bibitem [{\citenamefont {Gaveau}\ and\ \citenamefont
  {Schulman}(1997)}]{Gaveau1997}%
  \BibitemOpen
  \bibfield  {author} {\bibinfo {author} {\bibfnamefont {B.}~\bibnamefont
  {Gaveau}}\ and\ \bibinfo {author} {\bibfnamefont {L.}~\bibnamefont
  {Schulman}},\ }\href {\doibase https://doi.org/10.1016/S0375-9601(97)00185-0}
  {\bibfield  {journal} {\bibinfo  {journal} {Phys. Lett. A}\ }\textbf
  {\bibinfo {volume} {229}},\ \bibinfo {pages} {347 } (\bibinfo {year}
  {1997})}\BibitemShut {NoStop}%
\bibitem [{\citenamefont {Large}\ \emph {et~al.}(2018)\citenamefont {Large},
  \citenamefont {Chetrite},\ and\ \citenamefont {Sivak}}]{Large2018}%
  \BibitemOpen
  \bibfield  {author} {\bibinfo {author} {\bibfnamefont {S.~J.}\ \bibnamefont
  {Large}}, \bibinfo {author} {\bibfnamefont {R.}~\bibnamefont {Chetrite}}, \
  and\ \bibinfo {author} {\bibfnamefont {D.~A.}\ \bibnamefont {Sivak}},\ }\href
  {http://stacks.iop.org/0295-5075/124/i=2/a=20001} {\bibfield  {journal}
  {\bibinfo  {journal} {Europhys. Lett.}\ }\textbf {\bibinfo {volume} {124}},\
  \bibinfo {pages} {20001} (\bibinfo {year} {2018})}\BibitemShut {NoStop}%
\bibitem [{\citenamefont {Zulkowski}\ and\ \citenamefont
  {DeWeese}(2015)}]{zulkowskiNewMethod}%
  \BibitemOpen
  \bibfield  {author} {\bibinfo {author} {\bibfnamefont {P.~R.}\ \bibnamefont
  {Zulkowski}}\ and\ \bibinfo {author} {\bibfnamefont {M.~R.}\ \bibnamefont
  {DeWeese}},\ }\href {https://doi.org/10.1103/PhysRevE.92.032117} {\bibfield
  {journal} {\bibinfo  {journal} {Phys. Rev. E.}\ }\textbf {\bibinfo {volume}
  {92}},\ \bibinfo {pages} {032117} (\bibinfo {year} {2015})}\BibitemShut
  {NoStop}%
\bibitem [{\citenamefont {Tafoya}\ \emph {et~al.}(2018)\citenamefont {Tafoya},
  \citenamefont {Large}, \citenamefont {Liu}, \citenamefont {Bustamante},\ and\
  \citenamefont {Sivak}}]{TafoyaLarge2018}%
  \BibitemOpen
  \bibfield  {author} {\bibinfo {author} {\bibfnamefont {S.}~\bibnamefont
  {Tafoya}}, \bibinfo {author} {\bibfnamefont {S.}~\bibnamefont {Large}},
  \bibinfo {author} {\bibfnamefont {S.}~\bibnamefont {Liu}}, \bibinfo {author}
  {\bibfnamefont {C.}~\bibnamefont {Bustamante}}, \ and\ \bibinfo {author}
  {\bibfnamefont {D.}~\bibnamefont {Sivak}},\ }\href {\doibase 10.1101/291989}
  {\bibfield  {journal} {\bibinfo  {journal} {bioRxiv}\ } (\bibinfo {year}
  {2018}),\ 10.1101/291989}\BibitemShut {NoStop}%
\bibitem [{\citenamefont {Crooks}(2000)}]{HeatWork}%
  \BibitemOpen
  \bibfield  {author} {\bibinfo {author} {\bibfnamefont {G.~E.}\ \bibnamefont
  {Crooks}},\ }\href {https://doi.org/10.1103/PhysRevE.61.2361} {\bibfield
  {journal} {\bibinfo  {journal} {Phys. Rev. E.}\ }\textbf {\bibinfo {volume}
  {61}},\ \bibinfo {pages} {2361} (\bibinfo {year} {2000})}\BibitemShut
  {NoStop}%
\bibitem [{\citenamefont {Watanabe}\ \emph {et~al.}(2013)\citenamefont
  {Watanabe}, \citenamefont {Hayashi}, \citenamefont {Ueno},\ and\
  \citenamefont {Noji}}]{Watanabe2013}%
  \BibitemOpen
  \bibfield  {author} {\bibinfo {author} {\bibfnamefont {R.}~\bibnamefont
  {Watanabe}}, \bibinfo {author} {\bibfnamefont {K.}~\bibnamefont {Hayashi}},
  \bibinfo {author} {\bibfnamefont {H.}~\bibnamefont {Ueno}}, \ and\ \bibinfo
  {author} {\bibfnamefont {H.}~\bibnamefont {Noji}},\ }\href {\doibase
  10.1016/j.bpj.2013.09.050} {\bibfield  {journal} {\bibinfo  {journal}
  {Biophys. J.}\ }\textbf {\bibinfo {volume} {105}},\ \bibinfo {pages} {2385}
  (\bibinfo {year} {2013})}\BibitemShut {NoStop}%
\end{thebibliography}%
	
\end{document}